\documentclass[10pt,prx,aps,amssymb,twocolumn,superscriptaddress, notitlepage,longbibliography]{revtex4-2}

\usepackage{dcolumn}
\usepackage{bm}
\usepackage{braket}
\usepackage{mathtools}
\usepackage{ulem}
\usepackage[breaklinks,colorlinks=true,linkcolor=blue,urlcolor=blue,citecolor=blue]{hyperref}
\usepackage{times}
\usepackage{physics}
\usepackage{placeins}
\usepackage{latexsym}
\usepackage{amsmath, amssymb}
\usepackage{mathrsfs}
\usepackage{mathtools}
\usepackage{multirow}

\newcommand{\be}{\begin{eqnarray}}
\newcommand{\ee}{\end{eqnarray}}

\renewcommand{\theequation}{\arabic{equation}}

\begin{document}

\title{Microscopic study of topological phase transitions: Percolation point of view}
\date{\today}
\author{Keisuke Kataoka}
\affiliation{Department of General Education, Faculty of Science and Technology, Meijo University, Nagoya 468-8502, Japan}
\author{Ikuo Ichinose} 
\thanks{A professor emeritus}
\affiliation{Department of Applied Physics, Nagoya Institute of Technology, Nagoya 466-8555, Japan}

\begin{abstract} 
In this paper, we investigate how a topologically ordered state changes to another one from a microscopic
point of view.
To this end, we focus on the behavior of topological entanglement and 1-form symmetry under decoherence,
which induces a phase transition between these states.
After detailed numerical study of the color code as a first case study, 
we explain how the decoherence changes the topological properties of the state, observing topological entanglement negativity (TEN) 
and disorder parameter of the 1-form symmetry.
Our results suggest that the above quantities are closely connected with the first and 0-th simplicial homology, respectively, and that the transition can be understood from the point of view of decoherence percolation.
For detailed microscopic investigation, we introduce the concept of quasi-local TEN (QLTEN), 
and show snapshots of local topological property of the state, which is capable of clearly describing the decoherence percolation.
To further investigate how QLTEN and the 1-form symmetry are related to the transition, 
we introduce the concept of explosive percolation (EP) to the present system, which has received a lot of attention 
in the last two decades.
EP in the present context means that there exists some bias in the decoherence, reducing a large cluster of decohered regions.
As a second case study, we consider a toric code on a triangular lattice under an external-field-type decoherence.
Detailed numerical study reveals that QLTEN can faithfully describe the emergent Higgs region and 
the surviving logical qubits.
We find that although TEN, cluster of QLTEN and string operators strongly correlate with each other
in both the color code and the toric code, the two systems react to patterns of EP quite differently.
This fact indicates that obtaining a universal understanding of the topological phase transition
is challenging even for strongly-related systems.
\end{abstract}


\maketitle

\section{Introduction}

Needless to say, topological order is one of the most important concepts in physics and plays a significant role 
in quantum information science.
In this work, we shall study phase transitions from a topological state to another topological state
and also from a topological state to a non-topological state caused by quantum decoherence.
Topologically-ordered states have two distinct but strongly-correlated aspects, namely, 
quantum entanglement and 1-form symmetry.
Therefore, there are (at least) two indicators for describing the topological phase, such as the topological entanglement 
entropy (TEE) \cite{Levin2006,Kitaev2006} and 
(dis)order parameter of the 1-form symmetry \cite{Gaiotto_2015,mcgreevy2023}.
On the other hand, to the best of our knowledge, a full understanding of topological phase transition is still lacking,
such as how topological entanglement and 1-form symmetry are intertwined in the intermediate critical 
regime of the transition and whether indicators of the above two aspects
exhibit a singular behavior at the same point and with the same criticality, and if not, 
how this phenomenon should be understood.
 Clarifying these issues is an important step in deepening our understanding of the topological state.

The goal of this work is twofold.
First, we study the critical regime of the topological transition through two kinds of case study;
one from the color code to the toric code (TC) and the other from the TC to Higgs phase.
In the previous study \cite{kataoka2026measurement}, we investigated TC on a triangular lattice through the measurement-only circuit (MOC)
and found that the TEE and the 1-form symmetries exhibit rather distinct behavior in the critical regime, including
critical points and critical exponents.
We pursue this problem in this work.
To this end, we focus on a novel perspective and related physical quantities to elucidate the critical
regime in the microscopic view.
This may seem odd, as the topological property of the system is of a global nature.
However, physical measures observing the topological order such as TEE take an integer value
distinctive of each phase, and then a quasi-local quantity for topologically-ordered regions
can be defined, just as droplets of paramagnetic nature in the ordered phase of the Ising magnet.
More precisely, we introduce an observable for studying the critical regime of the above mentioned topological phase transitions,
that is, a measure to be regarded as quasi-local topological entanglement negativity (QLTEN), and by means of it, 
we obtain a rather clear physical picture of topological entanglement and 1-form symmetry.
There, view in terms of percolation and simplicial homology plays an essential role.

Second, motivated by observation on the data from the above perspective,
we introduce biased decoherence.
As explained in the above, viewpoint of percolation and network growth of decoherence is quite useful to understand 
the critical behavior of the system.
In the percolation study, phenomenon called explosive percolation (EP) has received a lot of attention in the last
two decades.
In the present context, decoherence with EP nature can be prescribed by employing a specific rule to 
determine the decoherence pattern, i.e., appearance of large clusters of decoherence is suppressed.

We study how the EP rule influences topological entanglement and 1-form symmetry, i.e.,
how topological states react to different decoherence patterns.
This is nothing but a problem of the {\it robustness of the topologically-ordered states against the biased decoherence 
in the critical region.}
We find that although topological entanglement negativity (TEN), cluster of QLTEN and 1-form symmetry correlate with each other
in both the color code and the TC, {\it the two systems exhibit quite different behaviors
under the EP decoherence.}


The paper is organized as follows. 
In Sec.~II, we introduce the color code, specific type of decoherence, and the stabilizer formalism
describing the target system.
In the stabilizer formalism, the target state is characterized by a set of Pauli strings
$\{g_\ell\}$ named stabilizer group, in such a way $g_\ell |\;\cdot\;\rangle=|\;\cdot\;\rangle$,
and the evolution of the state is observed by a change of the stabilizer generators,
an independent set of the stabilizer group.
The decoherence used here physically corresponds to the hopping of $X$-anyons on red links,
which induces the proliferation of $X$-anyons.
We clearly explain how the color code transits to the TC under the decoherence, in particular, 
how qubits on sites of a honeycomb lattice are transformed into qubits on links on a triangular lattice.
The emerging 1-form symmetry is also explained in detail, including a disorder parameter for that 
1-form symmetry.
We also introduce the TEN and the numerical formalism to calculate it.

In Sec.~III, we display numerical data for TEN and expectation value of the string operator as a function 
of the decoherence strength.
These data clearly show that the color code state changes to the TC on the emergent triangular lattice, as a result
of $X$-anyon proliferation.
However, a close look at the data indicates that there exists a discrepancy between the behavior of the entanglement entropy
and 1-form symmetry.
We qualitatively explain this discrepancy from a microscopic point of view.
There, perspective of percolation, network growth, and simplicial homology plays a very important role.
In particular, we introduce a microscopic version of TEN based on the perspective of simplicial homology, 
QLTEN.
We explicitly display snapshots of critical states obtained by calculating QLTEN in the entire system,
which provide a geometric overview of the transition under decoherence.

In Sec.~IV, we introduce biased decoherence, namely, EP that we explained above.
We propose various versions of EP and numerically investigate QLTEN and string operators.
We find that cluster formation observed by QLTEN is quite interesting from the viewpoint of percolation,
and this finding supports our observation obtained for the interrelation between topological entanglement and 1-form symmetry.

In Sec.~V, as a second case study, we investigate the TC, in particular, the relation with QLTEN and the logical 
qubits.
More precisely, we show that cluster formation of the Higgs region observed through QLTEN is a key point 
for fault tolerance of the TC.
The results obtained are relevant to the mitigation of quantum errors, which is one of the most actively investigated subjects
in quantum computing.

Section VI is devoted to the conclusion and discussion.
There, we emphasize that QLTEN and the other observables behave differently on the qualitative level in the above two case studies,
both of which are subject to the same EP decoherence. 
This fact indicates that obtaining a universal understanding of the topological phase transition
is challenging.
We hope that our work opens up several significant questions and research directions.

\section{Color code and emerging toric code}

In this section, we explain the target system and formalism for studying it
to make this paper as self-contained as possible, while
more details can be seen in our previous paper \cite{kataoka2026decohered}.

\subsection{Color code Hamiltonian and stabilizer formalism}
\begin{figure}[t]
\begin{center}
\includegraphics[width=8.5cm]{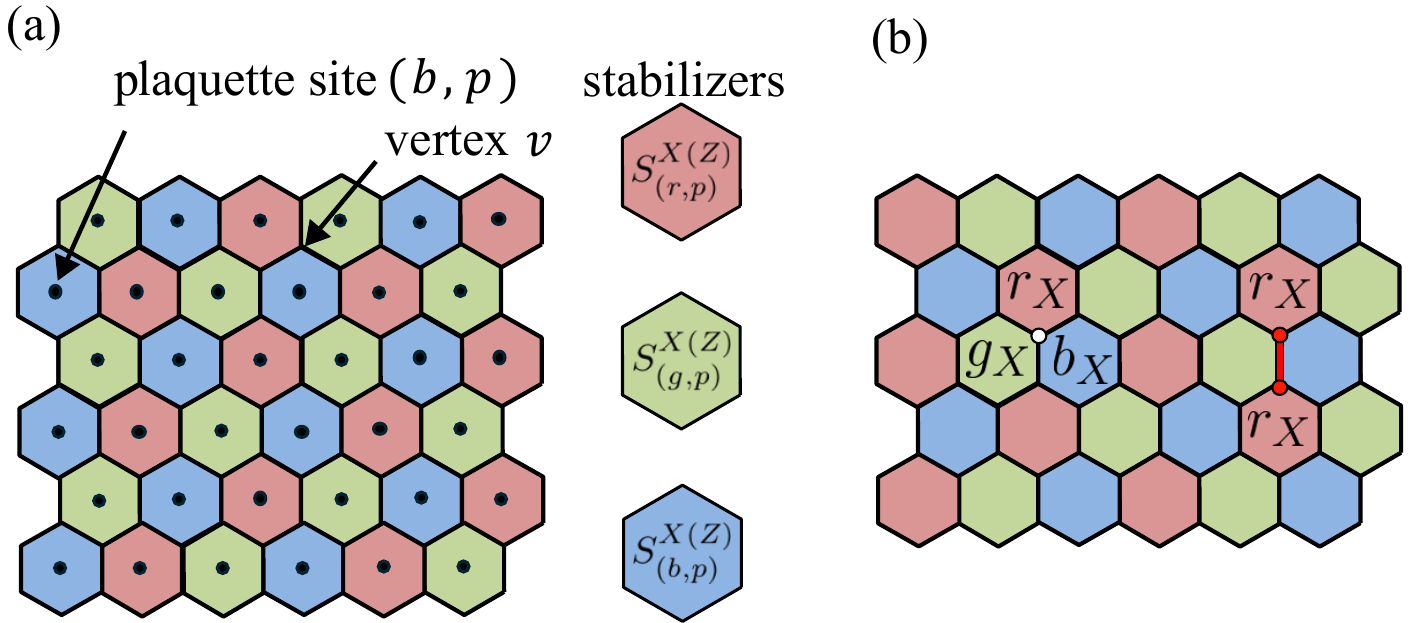}
\end{center}
\caption{(a) Schematics of the color code Hamiltonian.  
(b) By the operation of $X$ on a vertex produces three kinds of anyons in the 
vicinity plaquettes.
The operation of $XX$ on a red link produces a pair of red $X$ anyons.
}
\label{Fig_CCH}
\end{figure}

In this work, we study the transition from the color code to the TC \cite{Kitaev_1997,kitaev2003}, in particular,
critical states emerging by the decoherence effects.
In order to understand the properties of the color code, we look at the Hamiltonian that is given as follows
\cite{Bombin2006,Bombin2012,Bombin2015,Yoshida_2015,Kubica_2015},
\begin{eqnarray}
H_{\rm CC}=-\sum_{(p,c)}\biggl[S^X_{(c,p)}+S^Z_{(c,p)}\biggr],
\label{CC1}
\end{eqnarray}
\begin{eqnarray}
S^X_{(c,p)}\equiv \prod_{v\in (c,p)}X_v,\:\:\:\:\ S^Z_{(c,p)}\equiv \prod_{v\in (c,p)}Z_v,
\label{stab1}
\end{eqnarray}
where,$X_v$, $Z_v$ are Pauli operators residing on the vertex $v$ of the honeycomb lattice, 
$c$ refers to red($r$), green($g$) 
and blue($b$) ($c=r,g,b$), and $p$ labels a plaquette (hexagon). 
We use notations such that
$(c,p)$ refers to plaquettes and $v$ to six vertices around the plaquette $(c,p)$. 
The schematics of $H_{\rm CC}$ are shown in Fig.~\ref{Fig_CCH}(a).

Please note that all terms in $H_{\rm CC}$ commute with each other, and therefore 
the ground state $|{\rm GS}\rangle$ is given as
\begin{eqnarray}
S^X_{(c,p)}|{\rm GS}\rangle = S^Z_{(c,p)}|{\rm GS}\rangle = |{\rm GS}\rangle,
\label{GS1}
\end{eqnarray}
for all $(c,p)$.
The above facts mean that $S^{X(Z)}_{(c,p)}$ are stabilizers and 
the ground state $|{\rm GS}\rangle$ is a stabilizer state.
Excited states break at least one of the conditions in Eq.~(\ref{GS1}), and behave as anyon
as we explain later.
In the system subject to decoherence, such excitations are produced, and a change of the state
is traced by observing the stabilizer group.

On a torus, the stabilizers have redundancy such as \cite{Bombin2006}:
\begin{eqnarray}
\prod_{p}S^\alpha_{(r,p)}=\prod_{p}S^\alpha_{(g,p)}=\prod_{p}S^\alpha_{(b,p)},
\label{const1}
\end{eqnarray}
where $\alpha=X$ and $Z$. 
Then, the number of independent stabilizers called stabilizer generators is $N_v-4$, where $N_v$ is the total 
number of the vertex. 
Thus, the ground state degeneracy of the Hamiltonian $H_{\rm CC}$ is $2^{N_v-(N_v-4)}=2^4=16$,
meaning that the sector of the ground state encodes four logical (quantum) qubits~\cite{Bombin2006}.
From this fact, eight non-contractible logical operators can be constructed readily [not shown].
This topological aspect of the color code is quite similar to the TC.

From the above observation, the stabilizer formalism can be used for the color code 
with the following stabilizer set, $\mathcal{S}_{\rm CC}$,
\begin{eqnarray}
\mathcal{S}_{\rm CC}&=&\{S^X_{(r,p)}\}'+\{S^X_{(g,p)}\}'+\{S^X_{(b,p)}\}\nonumber\\
&+&\{S^Z_{(r,p)}\}'+\{S^Z_{(g,p)}\}'+\{S^Z_{(b,p)}\}
\equiv \{g_{\ell}\}.
\end{eqnarray}
There are four identities between the terms in $H_{\rm CC}$ as in Eq.~(\ref{const1}). 
Then, four arbitrarily chosen plaquette operators should be removed from 
the $X$-type and $Z$-type plaquette stabilizers. 
For example, $S^X_{(r,p)}$, $S^X_{(g,p)}$ and $S^Z_{(r,p)}$, $S^Z_{(g,p)}$ for specific plaquettes. 
The prime in $\{\cdot\}'$ indicates this manipulation and $\{g_{\ell}\}$ denotes the stabilizer generators
used with $\ell=0,1,\cdots, N_v-5$.

In terms of $\{g_{\ell}\}$, the density matrix of the ground state $\rho_{\rm CC}=|{\rm GS}\rangle\langle {\rm GS}|$
is expressed as 
\begin{eqnarray}
\rho_{\rm CC}\equiv \frac{1}{2^{4}}\prod^{N_v-5}_{\ell_=0} \biggr[\frac{I+g_{\ell}}{2}\biggr],
\label{DMCC}
\end{eqnarray}
where $I$ is the identity operator and, therefore, 
$g_\ell\rho_{\rm CC}=\rho_{\rm CC}g_\ell=\rho_{\rm CC}$ for all $\{g_\ell\}$.
[More precisely, we should explicitly denote the sector of the logical space.]

Let us consider how excited states, anyons, are described in the stabilizer formalism.
Anyons are created by some $X_v$ and $Z_v$ operators applied on the ground state $|{\rm GS}\rangle$.
As schematically shown in Fig.~\ref{Fig_CCH}(b), the $X_v$ operator on vertex $v$ excites three adjacent plaquettes
as it anti-commutes with the $Z$-plaquette operators.
In the stabilizer formalism, the state is expressed by the new stabilizer set $\{g'_\ell\}$, 
which includes $X_v$ and two composite operators out of three $Z$-plaquette operators
such as $S^Z_{(r,p_1)}\cdot S^Z_{(g,p_2)}$ and $S^Z_{(r,p_1)}\cdot S^Z_{(b,p_3)}$,
whereas the original three $Z$-plaquette operators are discarded.
Please note that these two composite operators and $X_v$ commute with each other.
The three emergent anyons are labeled as $(rX_v,gX_v,bX_v)$, respectively.

More interesting and important excitation is a pair of anyons.
Figure~\ref{Fig_CCH}(b) shows that a $XX$ operator on a red link, $X_{v}X_{v'}$, creates a pair of $X$ anyons in two red plaquettes $p_1,p_2$.
In the stabilizer formalism, the new stabilizer set includes $X_{v}X_{v'}$ and the composite operator such as 
$S^Z_{(r,p_1)}\cdot S^Z_{(r,p_2)}$, while the individual $S^Z_{(r,p_1)}$ and $S^Z_{(r,p_2)}$ are discarded.

The above manipulation of the stabilizer corresponds to the {\it measurement} of the operator $XX$ on a red link, 
in which the measurement results ($\pm 1$) are recorded and the state remains as a pure state.
In this work, we consider the effects of the $XX$-type {\it decoherence} on the color code without recording 
the measurement results.
In this setup, the $XX$ operator is {\it not} included in the stabilizer set and therefore the resultant state is mixed 
as a result of decoherence caused by interactions with environment, etc.

Measurement and decoherence drastically influence some important properties of the topological state such as 
the color code.
That is, the ground state of the color code, $|{\rm GS}\rangle$, is obviously invariant under the operation
of an arbitrary set of plaquette operators $\{S^{X/Z}_{(c,p)}\}$ and also a non-contractible loop of $X(Z)$ operators. 
These are nothing but 1-form symmetries.
In the stabilizer formalism, the 1-form symmetry is expressed as $\prod_{(c,p)}\Big[ S^X_{(c,p)}\Big]\rho_{\rm CC}=\rho_{\rm CC}$,
etc., which comes directly from Eq.~(\ref{DMCC}).
When anyons are generated by decoherence, e.g., interactions with environment, some of the plaquette operators are discarded
or deformed, and as a result, the original 1-form symmetry is reduced, although the topological order is robust.
In general setup of topological state, sufficiently dense decoherence (measurement) induces 
a phase transition from an original topologically-ordered state such as the TC and the color code to a less ordered state.
The study of this phenomenon is the subject of this work.

\subsection{Decoherence and emergent toric code}

\begin{figure}[t]
\begin{center}
\includegraphics[width=8cm]{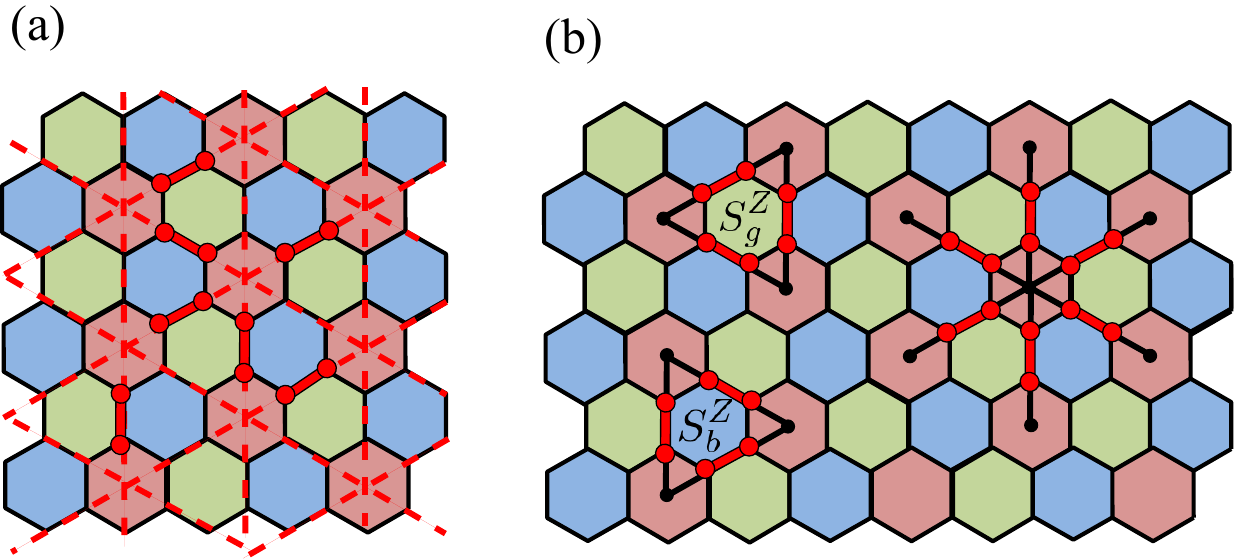}
\end{center}
\caption{(a) Schematics of emergent triangular lattice by red $XX$ decoherence 
denoted by thick line with dots at edges. 
Links connecting a pair of red plaquettes are called red links.
(b) Stabilizers of the emerging TC on the red triangular lattice.
}
\label{Fig_triangle}
\end{figure}
\begin{figure*}[t]
\begin{center}
\includegraphics[width=14cm]{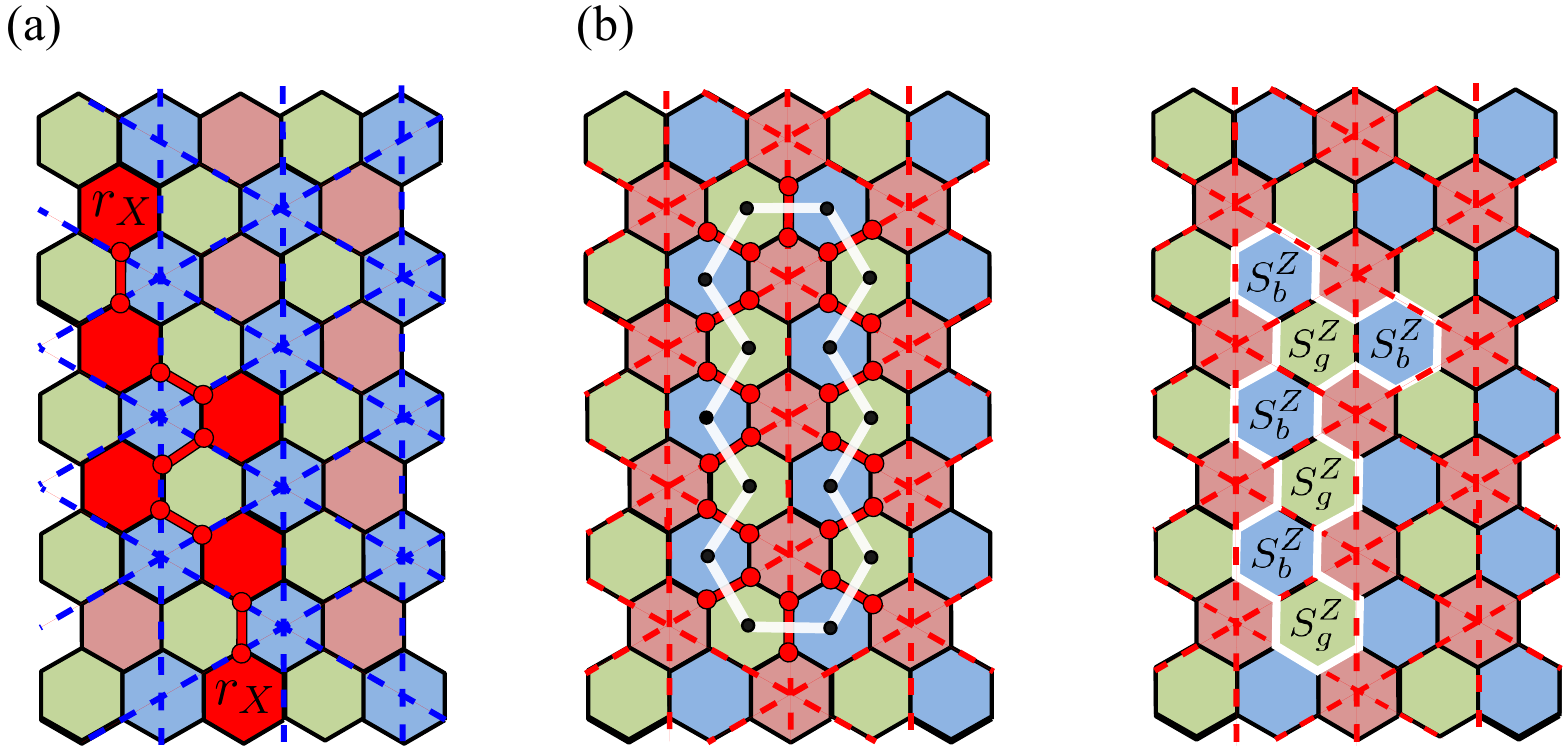}
\end{center}
\caption{(a) Schematics of $X$-anyon (so-called $m$-anyon in the TC) proliferation in the TC on {\it blue triangular lattice}
that loses quantum topological order.
(b) Emerging 1-form-symmetry operators of the TC on the {\it red triangular lattice}.
}
\label{Fig_CCtoTC}
\end{figure*}

In this subsection, we study the effect of $XX$-type decoherence applied on red links,
(red $XX$ decoherence).
Its explicit form is given in the stabilizer formalism as follows \cite{Nielsen2011},
\begin{eqnarray}
&&\mathcal{E}^{XX}=\prod_{(v_r,v'_r)}\mathcal{E}^{XX}_{(v_r,v'_r)},\nonumber\\
&&\mathcal{E}^{XX}_{(v_r,v'_r)}[\rho]=(1-p_r)\rho+p_r (X_{v_r}X_{v'_r})\rho (X_{v_r}X_{v'_r}),\nonumber\\
\label{deco_XX}
\end{eqnarray}
where $(v_r,v'_r)$ labels a red-link. 
In the present work, we stochastically set the parameter $p_r=1/2$ or $0$ for each red link 
with probability $p\in [0,1]$.
For $p=0$, the state is the original ground state and, with an increase of $p$, the state tends to mix.
In the limit $p=1$, the red $XX$ operator is applied to all red links.
This process can be regarded as {\it network growth} of the triangular lattice, vertices of which are red plaquettes.
See Fig.~\ref{Fig_triangle}(a).
From the point of view of anyons, the decoherence $\mathcal{E}^{XX}$ induces the proliferation of the ${\rm rX}$ 
anyon \cite{Wang_2025,Sohal2025}, the notion of which is different from condensation of anyons in pure states, 
since the proliferation of anyon cannot be regarded as a condensed object that is ``absorbed''
by the vacuum~\cite{Sohal2025}.

Let us closely look at the state emerging for the limit $p=1$, which is nothing but a TC state on the triangular lattice.
The $XX$-type decoherence on red link $(v_1,v_2)$ fixes the state on that link.
In the $Z$-diagonal representation, two kinds of emergent {\it link qubits} are 
\begin{equation}
\{|\uparrow\rangle_1|\uparrow\rangle_2\pm |\downarrow\rangle_1|\downarrow\rangle_2,
(|\uparrow\rangle_1|\downarrow\rangle_2\pm |\downarrow\rangle_1|\uparrow\rangle_2\},
\label{linkqubit}
\end{equation}
each of which corresponds to $XX=\pm 1$, as we are considering decoherence, not measurement.
It is easily verified that emergent link qubit operators are given as follows;
$\check{Z}_{v_1v_2}=Z_{v1}Z_{v2}$ and $\check{X}_{v_1v_2}=X_{v1}\mbox{ or } X_{v2}$, respectively.

It is known that the color code can be regarded as a product of two TC's, namely the color code 
$\sim$ (TC) $\times$ (TC) \cite{Bombin_2012NJP,Kubica_2015,Aloshious2019,Kesselring2024}.
For the $p=1$ case, the original qubits on all red links are applied by the $XX$ decoherence, and the remaining
degrees of freedom of the system are described by the link qubits in Eq.~(\ref{linkqubit})
and the link operators $\{\check{X}_\ell,\check{Z}_\ell\}$
that reside on the triangular lattice of the red links.
The color code stabilizers $S^Z_{(g,p)}$ and $S^Z_{(b,p)}$ tend to be stabilizers of the TC, 
$\prod_{\ell \in\triangle} \check{Z}_\ell$,
and also $S^X_{(r,p)}$ to $\prod_{\ell\in \check{v}}\check{X}_\ell$, where $\check{v}$ stands for the vertices of the triangular lattice. See Fig.~\ref{Fig_triangle}(b).
On the other hand, in the other TC, proliferation of the $X$-anyon takes place as shown in Fig.~\ref{Fig_CCtoTC}(a),
and it reduces that one part of TC in the color code to a trivial state having only classical qubit degrees of freedom.

For the TC on the red triangular lattice, two kinds of 1-form symmetries exist, as it should be.
The $Z$-type 1-form symmetry is given by a product of the emergent plaquette operators 
$\prod_{\ell \in\triangle} \check{Z}_\ell$, and the $X$-type 1-form symmetry by 
$\prod_{\ell\in \check{v}}\check{X}_\ell$, some of which are shown in Fig.~\ref{Fig_CCtoTC}(b).

\subsection{Basic observables for the color code to toric code transition}

For a transition of a topological state, there are two kinds of important properties to be observed,
namely topological entanglement and 1-form symmetry.
There are suitable observables for both of them; TEN and
the disorder parameter of the 1-form symmetry.
The stabilizer formalism is quite useful for calculating them as we explain below.

\subsubsection{Disorder parameter of 1-form symmetry}

We first consider the 1-form symmetry \cite{Gaiotto_2015,mcgreevy2023}, 
in particular the red $X$-1-form symmetry, whose charge is an arbitrary loop
composite of red $XX$ operators. 
Explicitly,
\be
W^{rX}(\gamma_c)\equiv \prod_{\ell\in \gamma_c} (XX)_\ell,
\label{WrX}
\ee
where $\gamma_c$ is a closed loop on red links and $(XX)_\ell$ stands for the $XX$ operator on the red link $\ell$.
It is not difficult to show that $W^{rX}(\gamma_c)$ can be expressed as the product of plaquette operators $S^X_{(g,p)}$ and 
$S^X_{(b,p)}$ inside the contractible loop $\gamma_c$.

Generally speaking, to see how symmetry is realized, namely if spontaneous symmetry breaking (SSB) 
takes place in the target state, disorder parameter is a very useful observable, in particular for 1-form symmetry.
For the symmetry whose charge operator is given by Eq.~(\ref{WrX}), the corresponding disorder parameter is given by
the expectation value of the following string operator as explained for the TC \cite{kuno2025intrinsic},
\be
{\cal D}^{rX}(\Gamma) \equiv \prod_{\ell\in \Gamma} (XX)_\ell,
\label{DrX}
\ee
where $\Gamma$ is an open string on red links, details of which are specified in the practical calculation.
Physically, the operator ${\cal D}^{rX}(\Gamma)$ is part of $W^{rX}(\gamma_c)$ and
creates a pair of rX-anyon at edges of the string $\Gamma$.
Therefore, it is obvious $\langle {\rm GS}|{\cal D}^{rX}(\Gamma)|{\rm GS}\rangle=0$ to indicate that
the rX-1-form symmetry is spontaneously broken in the ground state of the color code.
The TC has a similar property.

By the decoherence of Eq.~(\ref{deco_XX}), the ground state $|{\rm GS}\rangle$ tends to a mixed state,
\begin{eqnarray}
\rho_{\rm D}(p)\equiv\mathcal{E}^{XX}(\rho_{\rm CC}).
\label{rhoD}
\end{eqnarray}
For a mixed state, there are two different kinds of symmetries, strong and weak \cite{Buca_2012,Albert2014,groot2022}. 
For the symmetry in Eq.~(\ref{WrX}), the strong symmetry of a mixed state $\rho$ means 
$W^{rX}(\gamma_c)\rho=\rho W^{rX}(\gamma_c)=\rho$,
and the weak symmetry,
\begin{eqnarray}
W^{\rm rX}(\gamma_c)\mathcal\rho W^{{\rm rX}\dagger}(\gamma_c)=\rho.
\label{weaksymmetry}
\end{eqnarray}
In the stabilizer formalism, the strong symmetry means that the charge operator of the 1-form symmetry
is an element of the stabilizer group.
On the other hand, the weak symmetry is satisfied when the charge operator commutes with all elements
of the stabilizer group.
If spontaneous SSB occurs, it is examined by the disorder 
parameter~\cite{kuno2025intrinsic}.
In fact, the decohered state $\rho_{\rm D}(p)$ in Eq.~(\ref{rhoD}) can restore the weak 1-form symmetry
for a finite $p$~\cite{note1}.
Therefore, we focus on how the weak 1-form symmetry is realized in $\rho_{\rm D}(p)$.

As discussed in the previous work~\cite{kuno2025intrinsic}, the disorder parameter for weak symmetry is defined 
by using the disorder operator,
\be
D^X(\Gamma)=\mbox{Tr}[\rho_{\rm D}(p) {\cal D}^{rX}(\Gamma)\rho_{\rm D}(p){\cal D}^{rX}(\Gamma)]
/\mbox{Tr}[\rho^2_{\rm D}(p)].
\label{disorder2}
\ee
In the state with a finite value of $D^X(\Gamma)$, the weak 1-form symmetry is restored from the SSB.
Equation (\ref{disorder2}) indicates that symmetry restoration occurs as a result of proliferation of anyon.
This should be compared with the 1-form symmetry restoration in a pure state, in which anyon condensation plays an essential role.
In the following section, we show numerically that the restoration of the 1-form symmetry occurs in the weak-symmetry sense
as the parameter $p$ increases.

\begin{figure}[t]
\begin{center}
\includegraphics[width=8cm]{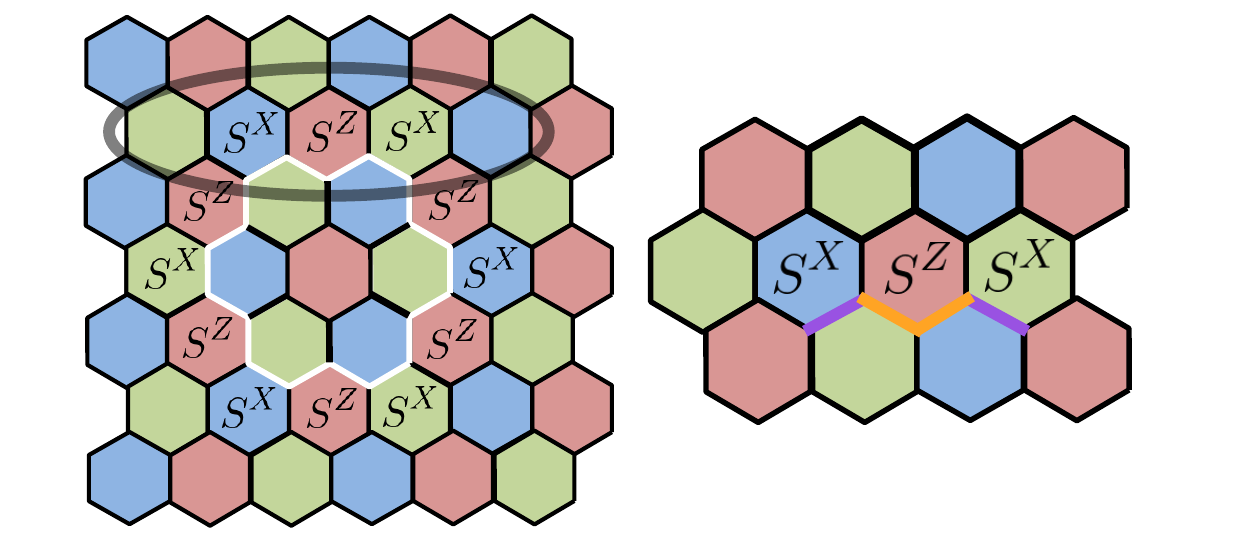}
\end{center}
\caption{Schematics of calculation of ${\cal N}_A$.
One type of $\tilde{K}_{1(2)A}$ is shown.
The other one is obtained by $X\leftrightarrow Z$.
Non-vanishing matrix elements come from the non-commutativity of truncated $S^{Z}$ and $S^X$
on the $A$-subsystem.
}
\label{Fig_CalTEN}
\end{figure}
\subsubsection{Topological entanglement negativity}

Moving on, let us discuss the quantum entanglement property of the system.
For pure systems, TEE is believed to be a good measure,
whereas it includes classical correlation as well for mixed states.
TEN was proposed as a measure of quantum entanglement, and recent studies indicate its utility.
We employ TEN for the present study.

Before introducing TEN, it is necessary to consider entanglement negativity or simply negativity, which 
is defined as follows for a chosen subsystem $A$,
\begin{eqnarray}
\mathcal{N}_A\equiv \log_2|\rho^{\Gamma_A}|_1=\ln|\rho^{\Gamma_A}|_1/\ln 2,
\end{eqnarray}
where $|\cdot|_1$ denotes the trace norm, and $\Gamma_A$ is a partial transpose of the basis in the subsystem $A$. 
This observable is a measure that can identify certain transitions among mixed states and their 
criticality~\cite{lu2020,sang2021,weinstein2022}.
In the stabilizer formalism, $\mathcal{N}_A$ can be efficiently calculated. 

$\mathcal{N}_A$ has a formula using the stabilizer generators $\{g_{\ell}\}$ such as 
\cite{sang2021,shi2021,sharma2022,KOI2023_1}. 
\begin{eqnarray}
\mathcal{N}_A=\frac{1}{2}\mathrm{rank}_{\mathbb{F}_2}\tilde{K}_A,
\label{N_A_formula}
\end{eqnarray}
where $\tilde{K}_A$ is a $m \times m$ matrix, $m$ is the total number of stabilizer generators of a state $\rho$,
$m \le N_v$. 
Explicitly, the matrix $\tilde{K}_A$ is obtained by searching the anti-commutative pairs for truncated stabilizer generators.
The truncated stabilizer generators are obtained in such a way that
the Pauli components of the generators residing within the complement of the subsystem $A$, $A^c$, are discarded. 
We denote them by $g^A_{\ell}$ ($\ell=0,1,\cdots, m-1$). 
The truncation procedure was explained in detail in the previous paper~\cite{kataoka2026decohered}.

\begin{figure*}[t]
\begin{center}
\includegraphics[width=13cm]{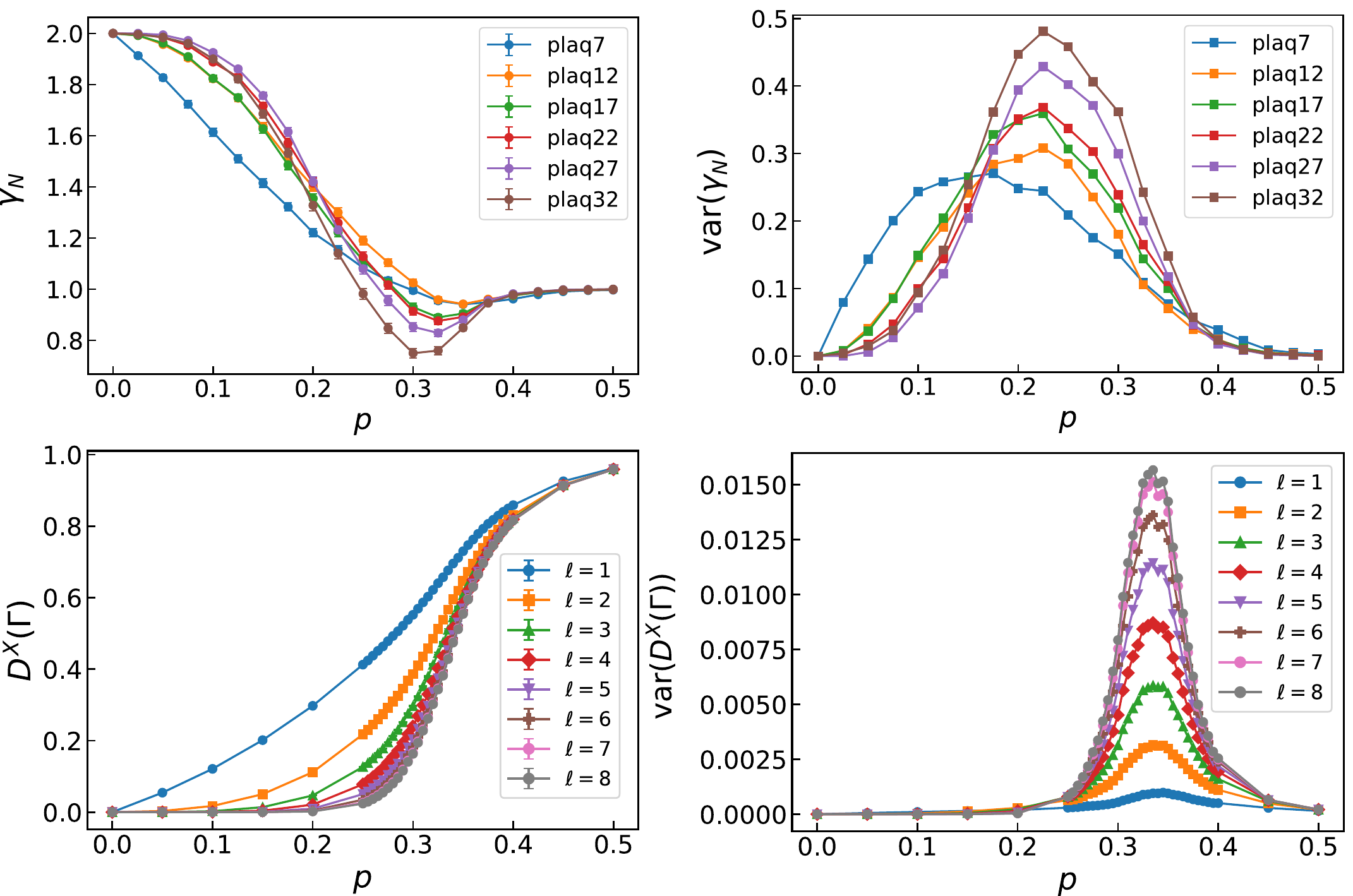}
\end{center}
\caption{Upper panels:$p$-dependence of topological entanglement negativity (TEN) and its variance.
Sizes of the subsystems $A,B,C$ are the number of plaquettes inside the subsystem.
Systematic system-size dependence is observed but the finite-size scaling analysis
cannot be applied in a satisfactory manner.
Lower panels:Expectation values of the string operator that work as a disorder parameter of
the 1-form symmetry.
$\ell=$ length of $\Gamma$.
The finite-size scaling analysis estimates the critical point $p_c=0.31$ and the critical exponent $\nu=1.2$.
The number of decoherence realizations is $10^3$ and the system size $(L_x,L_y)=(48,32)$.
}
\label{Fig_TEN1}
\end{figure*}

Using the truncated stabilizer generators $\{ g^A_{\ell}\}$, the matrix $\tilde{K}$ is defined as \cite{sang2021}
\begin{eqnarray}
(\tilde{K}_A)_{\ell,\ell'}=
\begin{cases}
1 & \mbox{if}\:\: \{g^{A}_{\ell},g^{A}_{\ell'}\}=0\\
0 & \mbox{if}\:\: [ g^{A}_{\ell},g^{A}_{\ell'}]=0
\end{cases},
\label{K_rule}
\end{eqnarray}
where $\{g^{A}_{\ell},g^{A}_{\ell'}\}$ ($[g^{A}_{\ell},g^{A}_{\ell'}]$) is anti-commutator (commutator).
It can be easily seen that the matrix $\tilde{K}$ is the binary matrix ($Z_2$) $m \times m$. 
The rank of the matrix $\tilde{K}$ exhibits the number of anti-commutative pairs of the truncated stabilizer 
generators. 
Pictorially, the stabilizer generators that touch the boundary of the subsystem $A$ are truncated, and as a result,
some of them tend to be anti-commutative with adjacent stabilizer generators. 
As the original stabilizer generators are $S^{Z(X)}_{(c,p)}$, $\tilde{K}$ has a symmetric matrix
structure, as anti-commuting property emerges through $X$ and $Z$ Pauli at the same site.
See Fig.~\ref{Fig_CalTEN}.

Recently, it was conjectured~\cite{Fan_2024} and verified by numerical calculation~\cite{kataoka2026decohered}
that the negativity $\mathcal{N}_A$ satisfies the following scaling law similar to that of the TEE,
\begin{eqnarray}
\mathcal{N}_A = c|\partial A|- \gamma_N + \cdots,
\label{NA_scaling}
\end{eqnarray}
where $c$ is a non-universal constant and $|\partial A|$ is the perimeter of the subsystem $A$. 
On the other hand, $\gamma_N$ is the TEN, which is universal in the sense that states belonging 
to an equivalent class of specific topological order have the same value of TEN. 
The term $c|\partial A|$ in ${\cal N}_A$ means that the state exhibits the area law of quantum entanglement.
More details of the law (\ref{NA_scaling}) will be explained in the following section.

The TEN can be extracted from the negativity of the subsystems by setting three adjacent subsystems, 
$A$, $B$ and $C$ such as
\begin{eqnarray}
\gamma_N &=& -\mathcal{N}_A - \mathcal{N}_B - \mathcal{N}_C - \mathcal{N}_{ABC}\nonumber\\
&+& \mathcal{N}_{AB} + \mathcal{N}_{BC} + \mathcal{N}_{AC}.
\label{def_TEN}
\end{eqnarray}
Throughout this work, for the value of TEN, the base of the logarithm is ``2'' since the negativity $\mathcal{N}_A$ is 
defined by using the same base of the logarithm.


\section{Numerical calculations:uniform decoherence}

In this section, we show the numerical calculations obtained by the stabilizer formalism.
As we explained earlier, we start with the color-code state and observe the effects of 
decoherence applied with probability $p\in [0,1]$ for each link.
This process can be regarded as some kind of network growth of decoherence
with {\it time step} $t\equiv pN_L$, where $N_L$ is the total number of red links of the honeycomb lattice 
and $t=0,1,2,\cdots, N_L$.
In this section, we employ the uniform decoherence configurations numbered $s$, and 
calculate the observables, ${\cal O}$, for each configuration sample to obtain $\langle {\cal O}\rangle_s$.
The mean value is given as 
\begin{eqnarray}
\langle \mathcal{O}\rangle = \frac{1}{N_s}\sum_{s}\langle \mathcal{O}\rangle_s, 
\label{average_observables}
\end{eqnarray}
where $N_s$ is the number of realizations of the decoherence.
In this section, we consider an unbiased ordinary decoherence and $N_s\sim 10^3$ in practical calculations. 


\subsection{Topological entanglement negativity}

We first study the topological negativity and the related TEN.
The subsystems used are shown in Fig.~\ref{Fig_TCABC} in Appendix A, 
and TEN is obtained by averaging the samples and locations of the subsystems.
For each setup, we show the numerical results in Fig.~\ref{Fig_TEN1}, in which the TEN changes from the value of 
the color code ($\gamma_N=2$) to that of the TC ($\gamma_N=1$).
The magnitude of the subsystems is defined with respect to the number of plaquettes inside of the subsystems.
We observe rather clear subsystem-dependence for the TEN and its variance, while the variance of the TEN is large compared with
the ordinary continuous phase transition consistent with the results in the previous paper \cite{kataoka2026decohered}, 
in which somewhat different subsystem shapes are used, as well as different boundary conditions.
We dared to apply the finite-size scaling (FSS) for the data of the variance of TEN, but have not obtained a satisfactory result.

It should be noted that the variance of the TEN exhibits a large and smooth peak at $p\simeq 0.22$,
and this value is obviously different from the bond percolation threshold on the triangular lattice $0.347$.
This result poses a question of how the system looks like in the `critical regime'.
We will answer this question after observing the behavior of the disorder parameter of 
the 1-form symmetry $D^X(\Gamma)$.

\begin{figure}[t]
\begin{center}
\includegraphics[width=6cm]{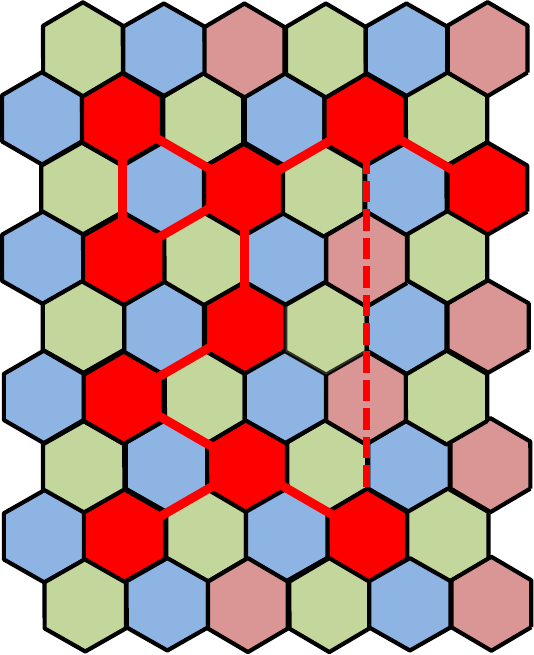}
\end{center}
\caption{Schematics of the disorder parameter of the 1-form symmetry; string operator (the dashed line).
For a non-vanishing value of the disorder parameter, two ends of the string have to be included in a 
single $S^Z$-cluster.
}
\label{Fig_pstring}
\end{figure}

\subsection{Disorder parameter of 1-form symmetry}

Let us turn to the disorder parameter $D^X(\Gamma)$.
The numerical data obtained are shown in Fig.~\ref{Fig_TEN1}, where the string $\Gamma$ is taken
as a straight line with the length measured as shown in Fig.~\ref{Fig_pstring}. 
As explained earlier, the finite value of $D^X(\Gamma)$ indicates that the string operator 
${\cal D}^{rX}$ in Eq.~(\ref{DrX}) commutes with all stabilizers at time step $pN_L$.
To realize this, the two original $S^Z$-plaquette stabilizers at the edges of the string $\Gamma$ have to be
merged into a large single cluster stabilizer by continuous decoherence, as shown in Fig.~\ref{Fig_pstring}.
The locations of the end points of $\Gamma$ are relevant, and this fact implies a close relationship between 
the 1-form symmetry and percolation, details of which will be discussed in the following subsection.

The data in Fig.~\ref{Fig_TEN1} show that $D^X(\Gamma)$ increases from 0 to 1 as $p$ increases for all $\Gamma$,
indicating rX anyon proliferation. 
This behavior clearly indicates that the transition from the color code to the TC occurs. 
The variance of $D^X(\Gamma)$ exhibits a sharp peak that indicates the continuous phase transition
with the critical point $p_s\simeq 0.31$ and the critical exponent $\nu\simeq 1.2$
obtained by the finite-size scaling. 
See Appendix A for details.
These values of the criticality are fairly close to those of the bond percolation point in the triangular lattice. 

The apparent discrepancy between the behavior of $\gamma_N$ and $D^X(\Gamma)$ poses a question, since it is believed that
quantum entanglement and 1-form symmetry generally arise from a single phase transition.
In the following subsection, we shall discuss this question from a microscopic perspective.
Roughly speaking, the large variance of TEN implies that there emerge a large variety of states
in the intermediate regime, which come from the non-trivial correlation between subsystem TENs
caused by its non-local definition.
This observation will be supported by investigating QLTEN in Sec.~IV.

 \subsection{Percolation and homology point of view}

\begin{figure}[t]
\begin{center}
\includegraphics[width=7.5cm]{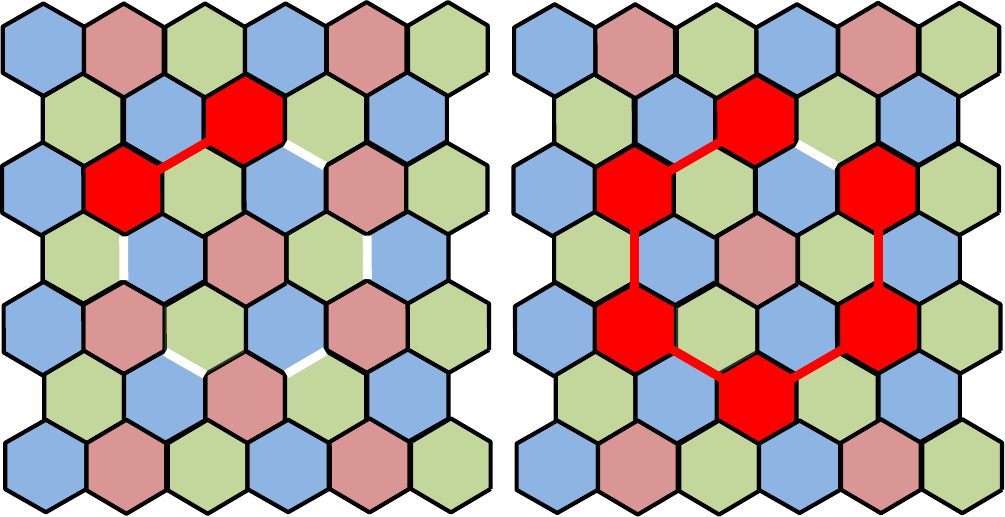}
\end{center}
\caption{Red $XX$ decoherence and anyons.
Left: Red $XX$ decoherence merges two red $S^Z$ stabilizers, generating
a pair of anyons.
Right: Typical quasi-closed loop of red $S^Z$ stabilizers, which is produced by a sequence of five
red $XX$ decoherence.
}
\label{Fig_anyonloop}
\end{figure}
\begin{figure}[t]
\begin{center}
\includegraphics[width=8.5cm]{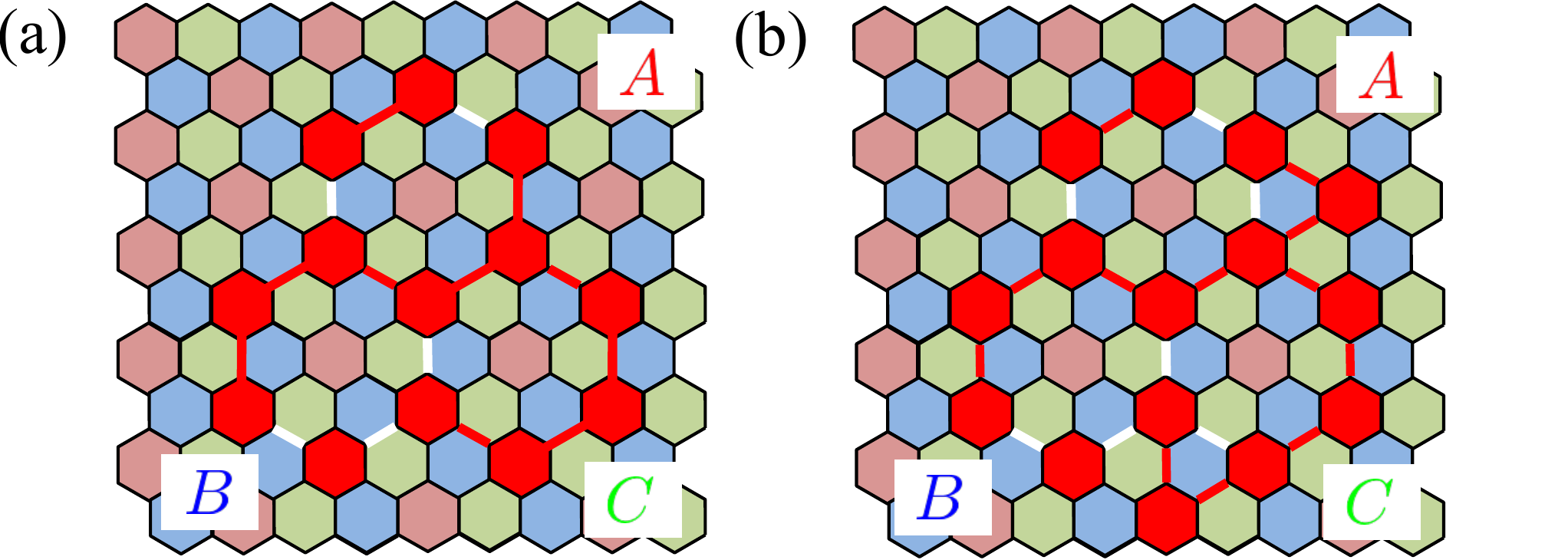}
\end{center}
\caption{Typical configurations of decoherence that have $\gamma_N=1$.
At least one of the $A,B,C$ subsystems is encompassed by a (quasi-)closed loop.
}
\label{Fig_confTEN}
\end{figure}

In the previous subsection, we explained how the 1-form symmetry is related to the bond percolation
on the triangular lattice.
In this subsection, we first elucidate the mechanism by which TEN $\gamma_N$ changes from 2 to 1
from the point of view of network growth.

\begin{figure*}[t]
\begin{center}
\includegraphics[width=14.5cm]{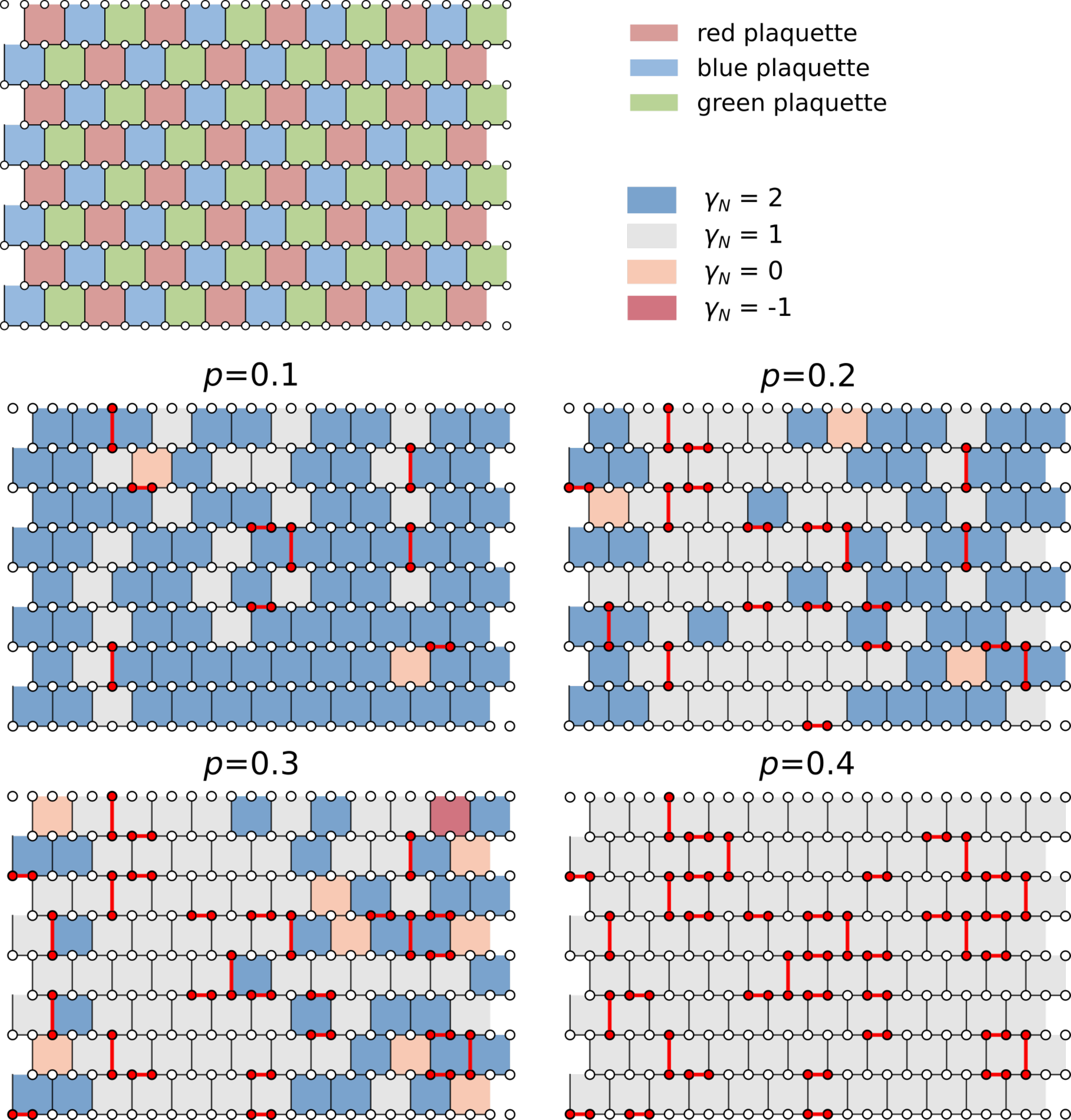}
\end{center}
\caption{Snapshots of QLTEN.
The top panel explains the map from honeycomb lattice to square lattice, which is used
in the display of QLTEN.
Snapshots clearly show how the TC regions evolve under decoherence.
At $p=0.4$, the whole system is in the TC region, which is consistent with the mean TEN in Fig.~\ref{Fig_TEN1}.
}
\label{Fig_CCsnap}
\end{figure*}

From the definition of the matrix $\tilde{K}_A$ [Eq.~(\ref{K_rule})], it has a block structure 
that is interchanged by $X\leftrightarrow Z$.
${\cal N}_A$ in Eq.~(\ref{N_A_formula}) has two independent contributions as can be schematically seen in
Fig~\ref{Fig_CalTEN}, and expressed as
\begin{eqnarray}
\mathcal{N}_A=\frac{1}{2}\big(\mathrm{rank}_{\mathbb{F}_2}\tilde{K}_{1A}+\mathrm{rank}_{\mathbb{F}_2}\tilde{K}_{2A}\big),
\label{N_A_formula2}
\end{eqnarray}
where the two contributions are interchanged by $X\leftrightarrow Z$.
Let us consider the specific type of subsystem whose center hexagon is red as in Fig.~\ref{Fig_CalTEN}, and then
by analytical calculation,
$\mathrm{rank}_{\mathbb{F}_2}\tilde{K}_{1A}=\mathrm{rank}_{\mathbb{F}_2}\tilde{K}_{2A}=|\partial A|-2$
for the color-coded ground state \cite{kataoka2026decohered}.
Here $|\partial A|$ is the number of stabilizers in $A^c$ that share the boundary of the subsystem $A$.
More precisely, the number of stabilizers $S^Z(S^X)=\frac{1}{2}|\partial A|$ and the number of red links
on the $A$ boundary is also $\frac{1}{2}|\partial A|$. 
For the 7-plaquette hexagon, $|\partial A|=12$.
See Fig.~\ref{Fig_CalTEN}.

A single red $XX$-decoherence application influences one of the block matrices, say $\tilde{K}_{2A}$, 
whose contribution to ${\cal N}_A$ decreases by 1, as numerically verified in Ref.~\cite{kataoka2026decohered}.
This comes from the following consideration, namely, application of $XX$ decoherence
at link $\ell$ connecting two red plaquettes induces the emergence of a pair of red $X$-anyon,
and is a network growth adding a bond at link $\ell$ of the red triangular lattice.
From Eq.~(\ref{K_rule}), this merging of two $S^Z$ stabilizers causes rank$\tilde{K}_{2A}$ to decrease by 2 and $N_A$ by 1.
Then, after operation of $\frac{1}{2}|\partial A|-1$ times along the $A$-boundary,
all $S^Z$-plaquette operators along the boundary have already merged into a large single stabilizer and 
therefore the remaining red $XX$ decoherence is redundant.
See Fig.~\ref{Fig_anyonloop}.
This picture is consistent with $\tilde{K}_{2A}-2(\frac{1}{2}|\partial A|-1)=0$.
Simply put, the anyon property of the $XX$-decoherence derives the scaling law Eq.~(\ref{NA_scaling}).

Let us turn to TEN.
By the above consideration and formula (\ref{def_TEN}), we can examine which configurations of the decoherence
reduce $\gamma_N=2 \to 1$.
We investigated various configurations and analytically calculated $\gamma_N$ composed of 7-plaquette subsystems $A,B,C$,
whose center hexagon is red in color.
See Fig.~\ref{Fig_TCABC}.
Then, we found that states, in which at least one of the three subsystems $A,B,C$ is (quasi-)closed, exhibit $\gamma_N=1$, 
where `(quasi-)closed subsystem' means that all $Z$-plaquette operators 
residing on the boundary of the subsystem are connected by decoherence effects.
See Fig.~\ref{Fig_confTEN} for examples in the network growth picture.
For the $\tilde{K}_{2}$ part, this is schematically expressed for the subsystems $A,B,C$
with the (quasi-)closed $C$ as follows;
\be
&& {\cal N}_A+{\cal N}_B+{\cal N}_C-{\cal N}_{AB}-{\cal N}_{BC}-{\cal N}_{CA}+{\cal N}_{ABC}   \nonumber \\
&& \Rightarrow {\cal N}_A+{\cal N}_B-{\cal N}_{AB}-{\cal N}_B-{\cal N}_A+{\cal N}_{AB}=0,\nonumber
\ee
while the contribution from the $\tilde{K}_{1}$ part remains intact.
General cases are rather difficult to examine analytically, but the above consideration 
provides us with a heuristic reflection on what is observed by numerical methods.

As explained earlier, the disorder parameter $D^X(\Gamma)$ searches for a connected cluster of  
$S^Z$-plaquette stabilizers including the end points of $\Gamma$ and TEN $\gamma_N$ 
for a (quasi-)closed sequence of $S^Z$-plaquette stabilizers, 
and therefore one may think that the 1-form symmetry and the topological entanglement are essentially the same thing.
However, this is not the case.
{\it From the point of view of simplicial homology, a cluster is a 0-th complex, whereas a closed loop is a first complex.}
This difference can be an origin of the distinct behaviors of $D^X(\Gamma)$ and $\gamma_N$ that are observed
numerically.

From the point of view of simplicial homology, the disorder parameter $D^X(\Gamma)$ works 
as a cohomology (that means a measure) of the 0-th homology.
One may wonder if there exists an operator to measure TEN from a microscopic perspective.
We propose the following operator as a cohomology of the (quasi-)closed loop;
\be
\prod_{\langle i,j\rangle\in \mbox{\small (quasi-)loop}}(\lambda + X_i X_j),
\label{quasiloop}
\ee
where $\langle i,j\rangle$ is an edge of a (quasi-)loop on the red triangular lattice and $\lambda$ is an arbitrary positive number.
Although we will not utilize this quantity in the present work, it will play an important role in 
the observation of the emergence of the $Z_2$ gauge theory property from the second homology perspective
by setting (quasi-)loops to triangles.
More comments are given in Sec.~VI.

As the perspective of homology is useful for understanding the topological transition, we are interested
in how the states in the intermediate regime look from the geometric point of view.
To this end, we introduce a measure, QLTEN, using 7-plaquette hexagons as subsystems $(A,B,C)$ 
and calculate their TEN by Eq.~(\ref{def_TEN}) as a quasi-local probe. 
In the configuration of values of QLTEN, we can find TC regions inside the color code phase
just like droplets in the ordered state of the Ising model.
The mean value of the QLTEN calculations is nothing more than $\gamma_N$ of the 7-plaquette subsystem, 
and then we are interested in the geometric distribution of QLTEN.
We note that QLTEN can be regarded as a cohomology for (quasi-)closed loop of the triangular lattice, whose
deep study might be an interesting mathematical issue.

Some snapshots of the QLTEN configurations are shown in Fig.~\ref{Fig_CCsnap}, in which the regions with $\gamma_N=1$ are
called TC region.
We find that the snapshots are quite instructive and provide us with an intuitive picture of the intermediate states,
which is beyond the previous studies on topological phase transition. 
In the following section, systematic investigation of the topological and entanglement properties of the system
is performed.
To this end, we introduce the concept of EP in the present study, which plays an important role in the study of percolation 
over the last two decades.
The study using EP can be seen as a thought experiment to obtain a deep understanding of topological order
and quantum entanglement, but the results obtained are useful for quantum error mitigation.

\begin{figure*}[t]
\begin{center}
\includegraphics[width=13cm]{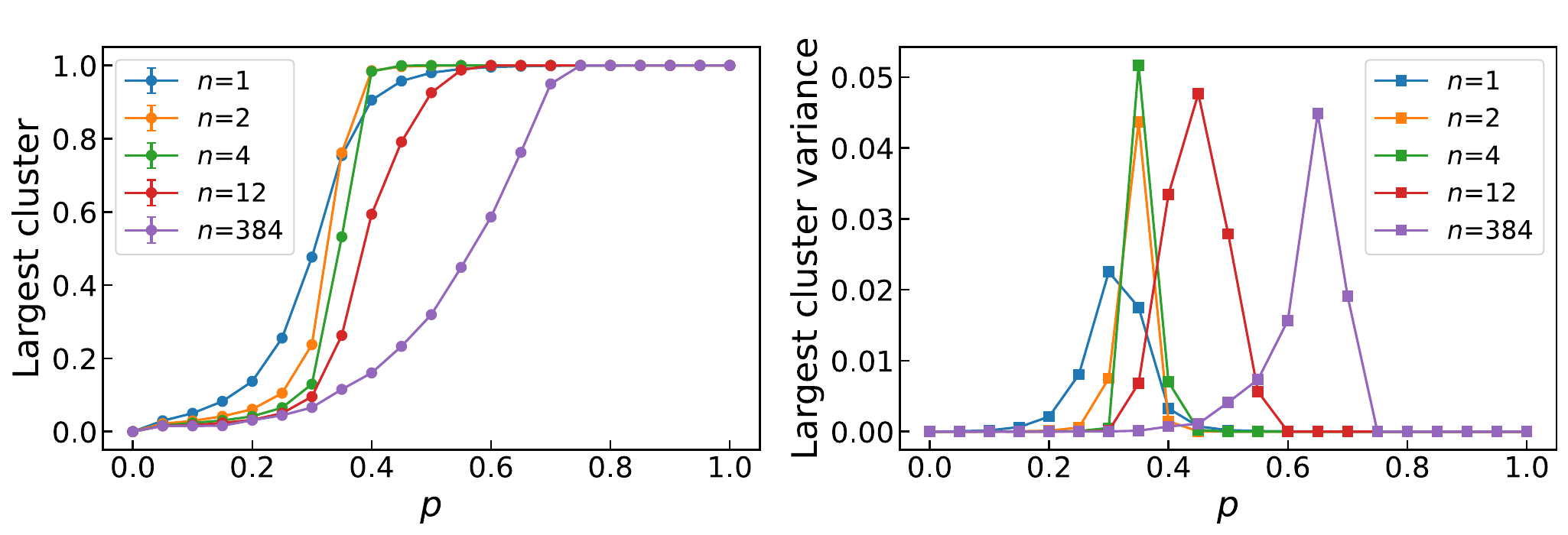}
\end{center}
\caption{ Largest clusters for various $n$-trial EP created by Union Find.
They exhibit clear $n$-dependence that plays an important role for study of topological transition.
The $n=384$ case is the maximal EP, in which all empty links are a candidate for decoherence applied on.
}
\label{Fig_EP}
\end{figure*}
\begin{figure*}[t]
\begin{center}
\includegraphics[width=13cm]{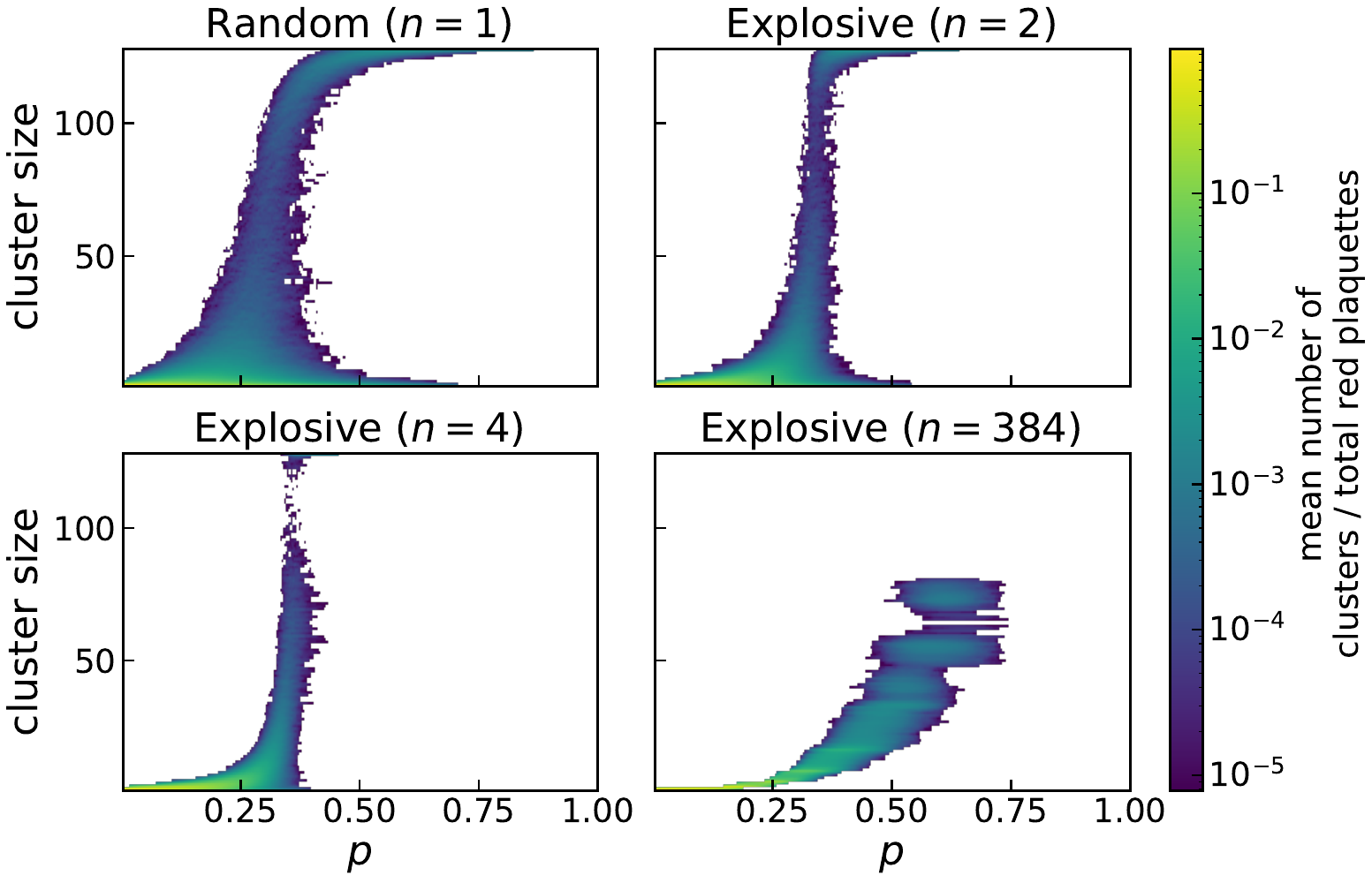}
\end{center}
\caption{Cluster size distributions of decoherence for various EP.
For $n=1$, the distribution changes rather smoothly as $p$ increases.
As $n$ is getting larger, the distribution is getting to exhibit a step-function like behavior, and
for very large $n$, the sizes of the clusters are limited, and a extremely large cluster suddenly emerges
at the critical point.
This behavior of decoherence clusters gives us an important insight into fault tolerance of the logical qubits
in topological codes.
See later discussion.
}
\label{Fig_clustersize}
\end{figure*}

\section{Numerical calculations: Biased decoherence}

\subsection{Explosive percolation}

In this section, we first explain EP \cite{li2024explosive} and the motivation to consider it in the present study.
EP refers to a modified bond percolation method often performed by the Achlioptas process \cite{achlioptas2009explosive}.
Standard bond percolation inserts bonds at random and independently.
The Achlioptas process selects bonds via rules that can suppress the growth of large clusters. 
In the present work, we employ the product rule.
Starting from a lattice with all empty bonds (the ground state of the color code),
at each time step, $n$ potential bonds are chosen randomly out of the empty bonds.
The sizes of the two clusters that attach to each bond are calculated
(if no cluster attaches to an end of bond, the size of `cluster'=1),
and the one with the smallest size product of the associated clusters is inserted.
We call this process $n$-trial EP, in which the insertion of the bond corresponds to the application of
red $XX$ decoherence for that bond.
For an emergent bond pattern at each time step, we calculate the QLTEN values for the whole system.
Obviously, 1-trial EP is the standard one. 

In addition to EP, we introduce an additional measure to determine the percolation threshold, which is nowadays 
standard and gives satisfactory finite-size scaling with respect to the system size.
After producing the QLTEN configurations using the EP bond algorithm by Union Find, 
the measure named event-based ensemble monitors the incremental one-step size of the largest TC cluster, defined as
$$
\Delta(t)=C_1(t+1)-C_1(t),
$$
where $C_1(t)$ is the size of the largest cluster of the TC region.
As the time step progresses, $\Delta(t)$ generally increases first, reaches a peak, and then starts to decrease.
For each decoherence process, the time $t$ for the largest $\Delta(t)$ determines the pseudo-critical time $t_{\mbox{\small max}}$.
Obviously $t_{\mbox{\small max}}$ gives a pseudo-critical threshold in the sense of decoherence event-bases, namely,
${p}_{\mbox{\small max}}=t_{\mbox{\small max}}/N_L$.
We shall examine the distribution of $t_{\mbox{\small max}}$ in a later investigation and see how it correlates
with the used bond distribution.
To this end, we utilize the Union-Find algorithm.

EP provides us with an extra-ordinary set up of decoherence.
It is quite interesting to see how QLTEN and 1-form symmetry change and which quantities 
correlate with each other under EP decoherence.
This observation sheds light on the microscopic properties of the topological phase transition.
In fact, as we shall see, the behavior of QLTEN depends on the system under consideration,
and therefore, QLTEN may be used as a fingerprint of phase transitions.

Figure~\ref{Fig_EP} displays how large clusters of the percolated region evolve as $p$ increases in various EP processes, which are
produced using Union Find for the triangular lattice.
Although the threshold is less sensitive to the percolation method (i.e., the value of $n$) than expected, 
a sudden increase in cluster size is observed for larger $n$.
These numerical results are consistent with those for the EP in the square lattice \cite{li2024explosive}.
The obtained behavior will be used in a later discussion on topological transition.

The numerically obtained values $\langle p_{\mbox{\small max}}\rangle$ are the following;
for $n=1$, $\langle p_{\mbox{\small max}}\rangle \simeq 0.320$,
for $n=4$, $\langle p_{\mbox{\small max}}\rangle \simeq 0.365$, and 
for $n=384$, $\langle p_{\mbox{\small max}}\rangle \simeq 0.652$.
These values strongly correlate with the location of peaks of the largest-cluster-size variance
shown in Fig.~\ref{Fig_EP}.


\begin{figure*}[t]
\begin{center}
\includegraphics[width=13cm]{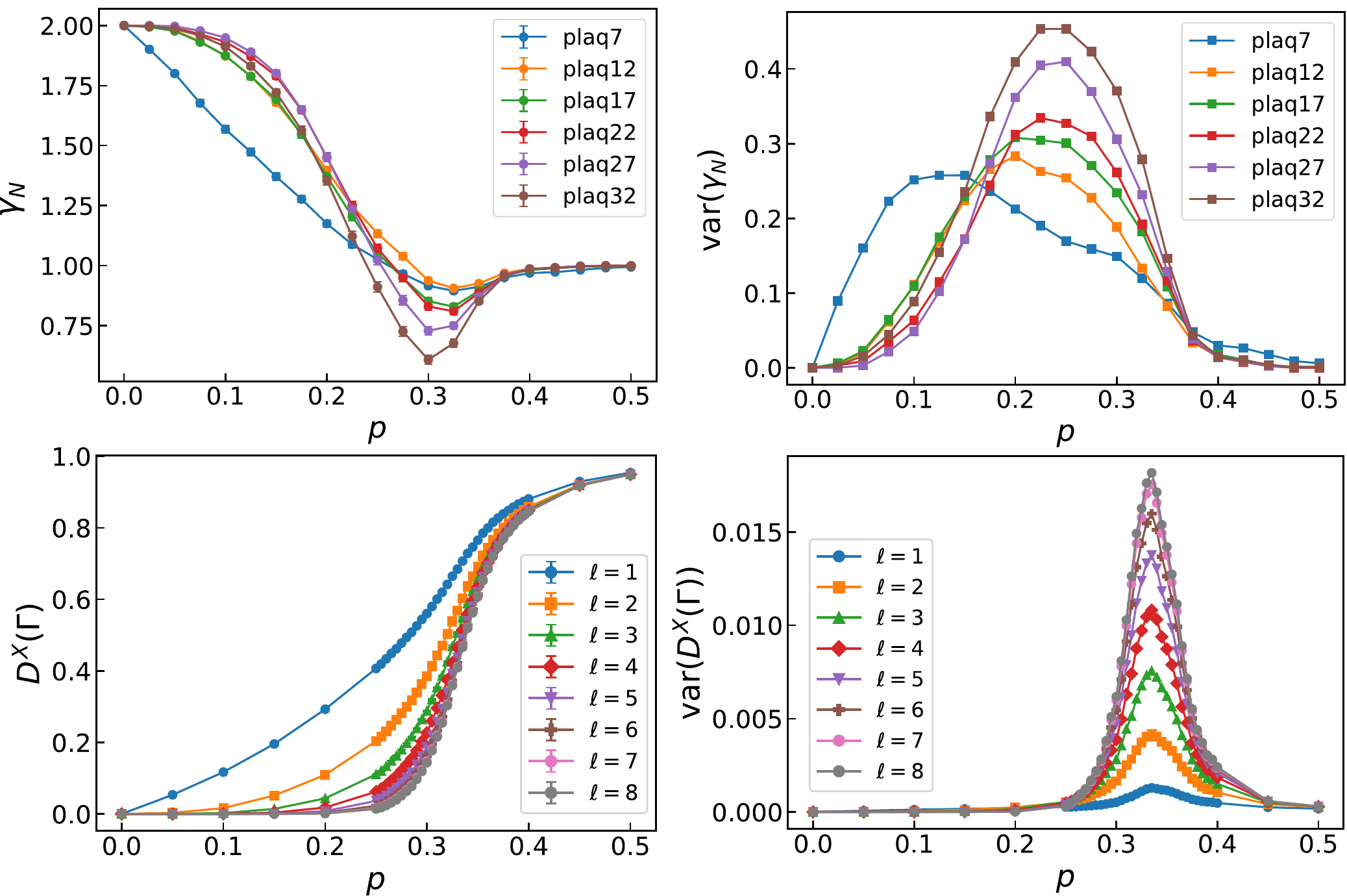}
\end{center}
\caption{Upper panels:TENs and their variances for $n=384$ EP.
Lower panels:Expectation values of the string operator and their variance.
These observables show only small $n$-dependence, see Fig.~\ref{Fig_TEN1} for $n=1$.
System size $(L_x,L_y)=(48,32)$.
}
\label{Fig_largen}
\end{figure*}

\subsection{Cluster size distribution obtained by simulation using Union Find}

Let us finally consider the cluster size  distribution obtained by the simulation using Union Find.
To the best of our knowledge, this study has not been done yet for the triangular lattice system,
and therefore, we report the simulation results rather in detail.

The cluster size distribution in Fig.~\ref{Fig_clustersize} provides us with 
important information about the critical regime of the topological transition.
That is, we shall later examine the evolution of the QLTEN cluster and compare it with the data of Union Find 
of the system subject to biased decoherence.
For EP of $n=1$, the distribution is rather smooth as a function of $p$, and for EP of $n=4$, 
the distribution exhibits a discontinuous jump at $p\simeq 0.35\simeq \langle t_{\mbox{\small max}}\rangle$.
For larger values of $n$, only small clusters emerge in a large amount and merge into a single large cluster
of the system size at $p\simeq 0.75$.
This behavior of clusters is closely related with fault tolerance of topological qubits as we discuss in the context
of the TC in the following section.

Another remark obtained from the data on the cluster size distribution is that 
there are a large number of clusters in the intermediate values of $p$.
This fact explains the behavior of the variance of the mean TEN in that region; the variance exhibits
a large but smooth peak as shown in Fig.~\ref{Fig_TEN1}.
More comments will be given after showing the numerical data for a large value of $n$ in
Fig.~\ref{Fig_largen}.


\begin{figure*}[t]
\begin{center}
\includegraphics[width=12cm]{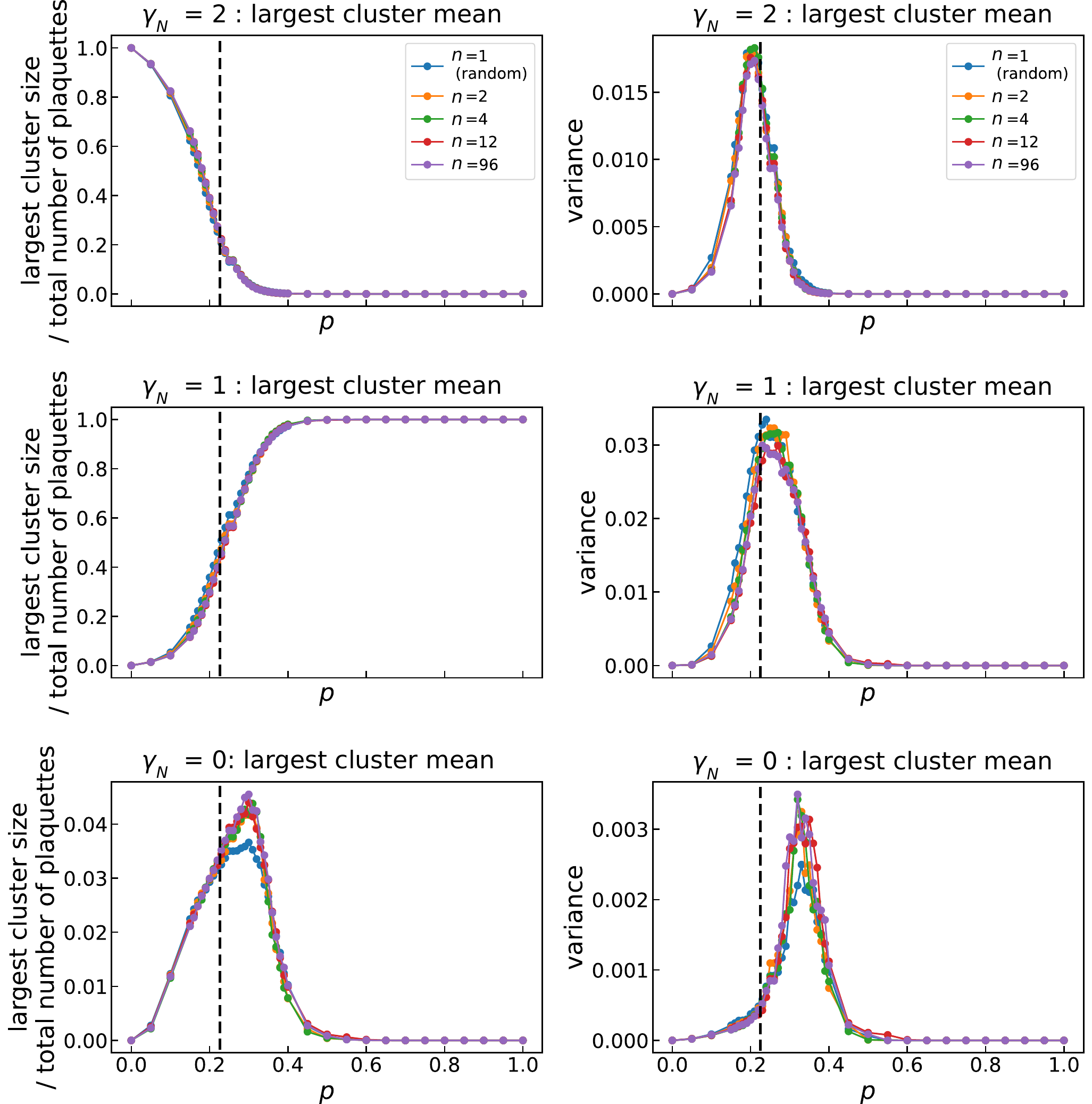}
\end{center}
\caption{Largest cluster sizes of the color code as a function of $p$, and
$n=1\sim 96$.
We observe very small EP-type dependence for the color code.
The dotted lines indicate the location of the transition $p_{\rm QL}=0.22(0.22)$ for $n=1(96)$
estimated through the event-based ensemble.
The obtained values have very small $n$-dependence.
System size $(L_x,L_y)=(12,8)$.
}
\label{Fig_CCcluster}
\end{figure*}

\subsection{Topological entanglement negativity and string operator for EP decoherence}

In this subsection, we numerically study the mean TEN and the string operator for $n$-trial EP decoherence
explained earlier in this section.
By practical calculations, we found that these observables have very small $n$-dependence, and therefore,
we show only the data for a very large $n$ case.

Figure~\ref{Fig_largen} shows the numerical calculations of the mean TEN and $D^X(\Gamma)$.
TEN data exhibit a very close size-dependence to that
of ordinary percolation $n=1$ shown in Fig.~\ref{Fig_TEN1}.
The disorder parameter $D^X(\Gamma)$ of the 1-form symmetry also has only a small dependence on $n$ for all $\Gamma$.
The critical value and the critical exponent obtained are $p_s\sim 0.31, \nu=1.2$ that are
very close to those of the $n=1$ EP.
For details, see Appendix A.

The above numerical study, in particular, on the mean TEN, implies that a more microscopic
investigation of the topological transition and the critical states is needed to understand
what is happening in the intermediate regime under EP decoherence.
This is a subject of the next subsystem.


\subsection{Largest cluster and threshold by QLTEN}

In this subsection, we show the behavior of the largest cluster and its variance, by which
we estimate the threshold of the phase transition.
Then, the data of the largest cluster are compared with those of the Union Find used to generate biased decoherence
in Fig.~\ref{Fig_EP}.

In Fig.~\ref{Fig_CCcluster}, we show the average size of the largest clusters of the TC region as well as the color code region,
and their variance as a function of $p$ for typical types of EP decoherence.
From the data obtained for the TC region, we see the unexpectedly small $n$-dependence 
without resemblance to the corresponding
quantities of EP on the triangular lattice shown in Fig.~\ref{Fig_EP}.
The location of the phase transition obtained by the event-based ensemble
is $p_{\rm QL}= \langle p_{\rm max}\rangle=0.22(0.22)$ for $n=1(96)$ in Fig.~\ref{Fig_CCcluster}, while we observed
$p_{\rm UF}=0.29(0.54)$ for $n=1(96)$.
We note that the value $p_{\rm QL}=0.22(0.22)$ for $n=1(96)$ is very close to the location of the peak
of the TEN variance for plaq 12 $\sim$ plaq 32 in Figs.~\ref{Fig_TEN1} and \ref{Fig_largen}, although 
QLTEN is obtained from plaq 7.

The above observation is an important finding of this work; 
In the color code system, the global values of TEN and the disordered parameter are insensitive to the decoherence property, 
and the clustering property of TC regions viewed via QLTEN is also insensitive.
One may think that the origin of this insensitivity is the topological order.
That is, {\it this behavior of the color code seems to indicate the robustness of its topological property
even in the phase-transition regime;
Whether this phenomenon is universal or depends on the details of the model system is a substantial issue.}
This issue will be clarified by the study on the TC, which is given in Sec.~V.


\begin{figure}[t]
\begin{center}
\includegraphics[width=8cm]{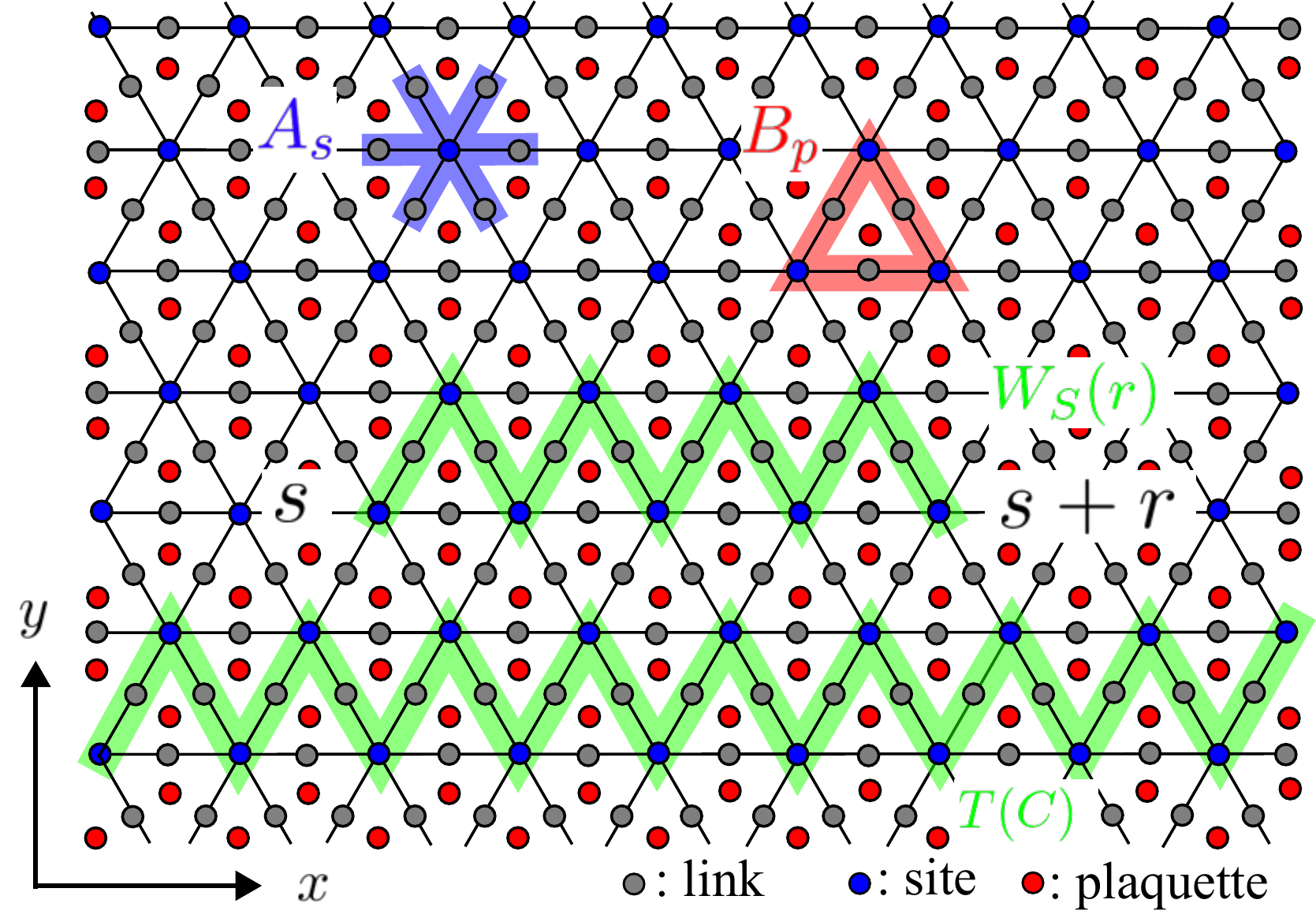}
\end{center}
\caption{Schematics of the toric code Hamiltonian. Wilson string and 'tHooft loop use similar zigzag line/loop
on the triangular lattice for the symmetry. 
}
\label{Fig_TCH}
\end{figure}

\begin{figure*}[t]
\begin{center}
\includegraphics[width=14cm]{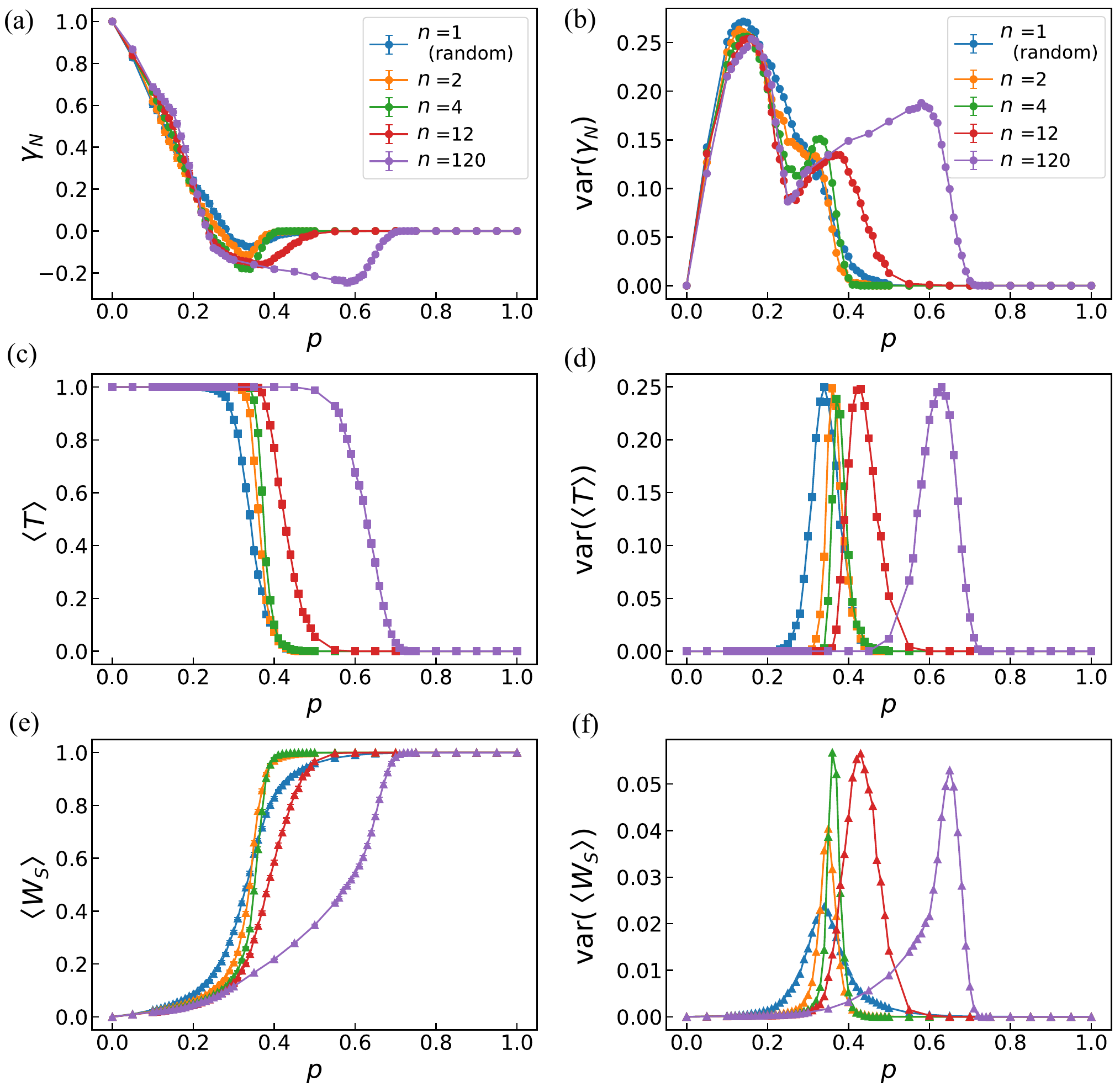}
\end{center}
\caption{Various observables of the toric code under various types of EP decoherence.
(a) and (b): TEN and its variance, which have very specific $p$-dependence for larger $n$-EP.
(c)-(f): Existence probability of the non-contractible $X$ and expectation value of $Z$ string, which show closely-correlated
behavior with $n$-EP in Fig.~\ref{Fig_EP}. 
System size $(L_x,L_y)=(12,12)$.
}
\label{Fig_TC1}
\end{figure*}
\begin{figure*}[t]
\begin{center}
\includegraphics[width=14cm]{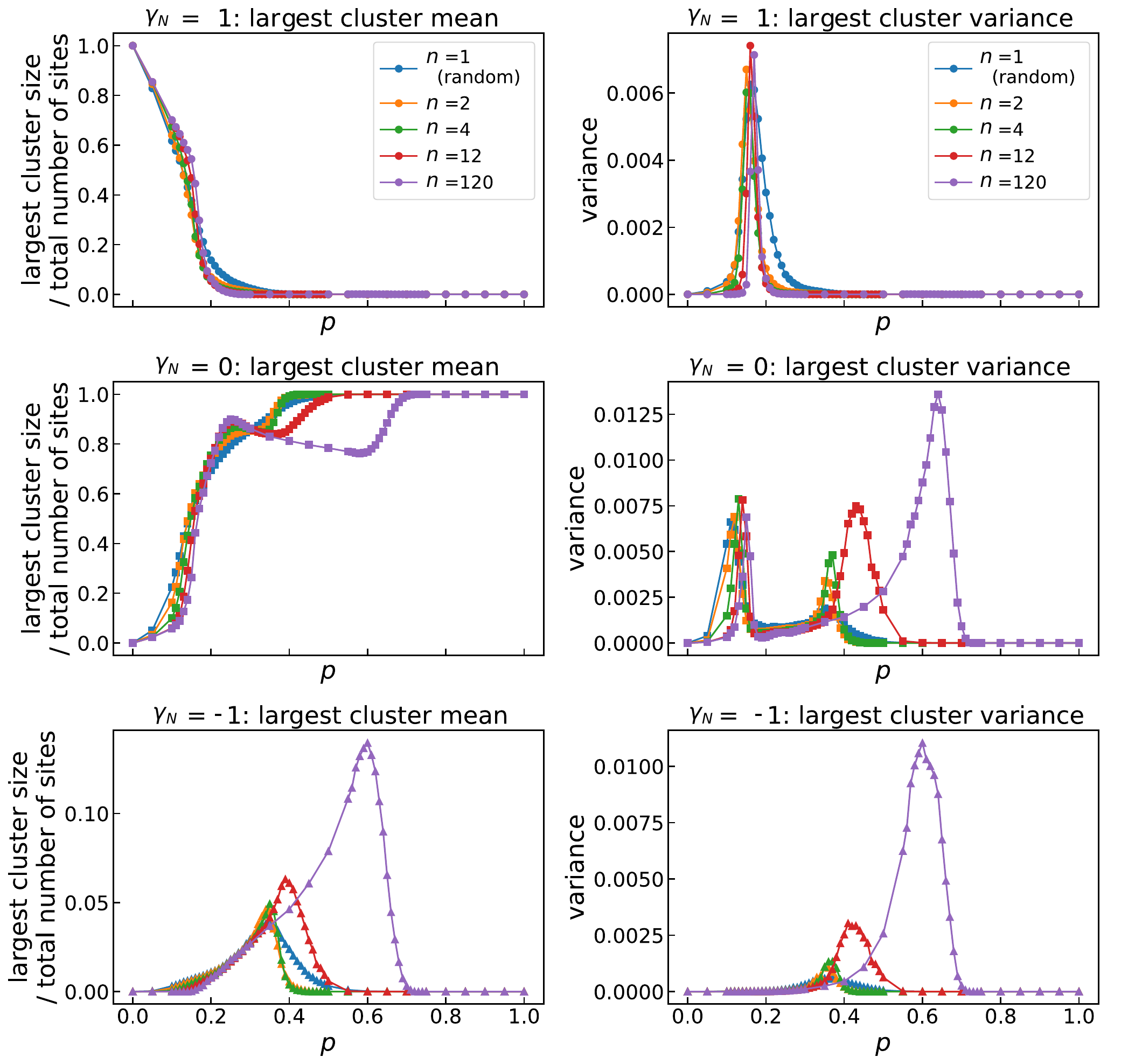}
\end{center}
\caption{Largest clusters of regions with $\gamma_N=1,0, -1$ observed by QLTEN and their variance.
System size $(L_x,L_y)=(12,12)$.
}
\label{Fig_cluster/TC}
\end{figure*}
\section{Toric code under decoherence on a triangular lattice}

In the previous sections, we studied the transition from the color code to the TC and found that
QLTEN plays an important role in describing the microscopic behavior of the system.
However, connected clusters emerging from the QLTEN configurations are insensitive
to the type of decoherence, similarly to the mean of TEN and the disorder parameter of the 1-form symmetry.
It is an important issue if this behavior is generic or model-dependent, which we address in this section.

We also note that there is a missing link, that is, how QLTEN is related to logical operators.
This issue is quite important from the point of view of quantum error mitigation in topological
quantum computing, which can be investigated by using methods similar to those used to minimize spreading epidemics
as we show now.
We tried to solve this issue in the color code context, but found that the TC is more suitable for that,
as qubits of the TC reside on links and the logical operators are a non-contractible continuous loop operator
defined without ambiguity.
Furthermore, in our previous paper \cite{kataoka2026measurement}, 
we investigated the behavior of non-contractible loop operators in the framework of
MOC and discussed its relationship with topological entanglement entropy.

In this section, we study the TC \cite{Kitaev_1997,kitaev2003} subject to decoherence as a second case study.
As qubits of the TC reside on links instead of vertices, 
physical observables may exhibit behavior different from those of the color code.
In the practical calculation, we use the decoherence corresponding to the $Z$ external field,
which geometrically corresponds to the $XX$-decoherence employed for the study of the color code,
and the QLTEN subsystems are taken as a hexagon as in the color code, as well as in the previous work \cite{kataoka2026measurement}. 
See Fig.~\ref{Fig_TCABC}.

\subsection{Toric code Hamiltonian and decoherence}

\begin{figure*}[t]
\begin{center}
\includegraphics[width=16cm]{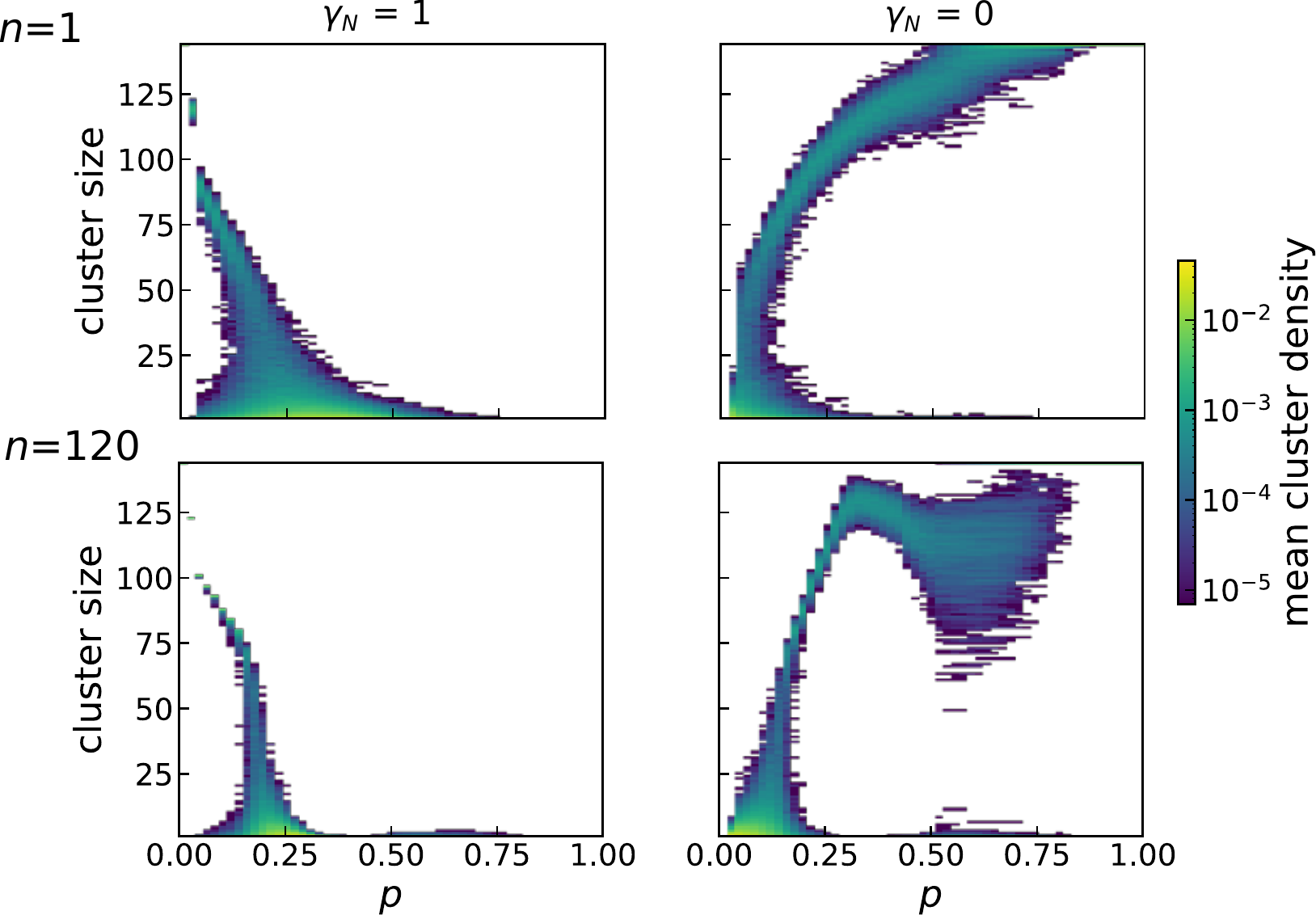}
\end{center}
\vspace{-0.5 cm}
\caption{Cluster-size distributions of TC ($\gamma_N=1$) and Higgs ($\gamma_N=0)$ regions 
in the toric code as a function of $p$ for $n=1$ and $120$.
System size $(L_x,L_y)=(12,12)$.
}
\label{Fig_TCdistribution}
\end{figure*}

\begin{figure*}[t]
\begin{center}
\includegraphics[width=14cm]{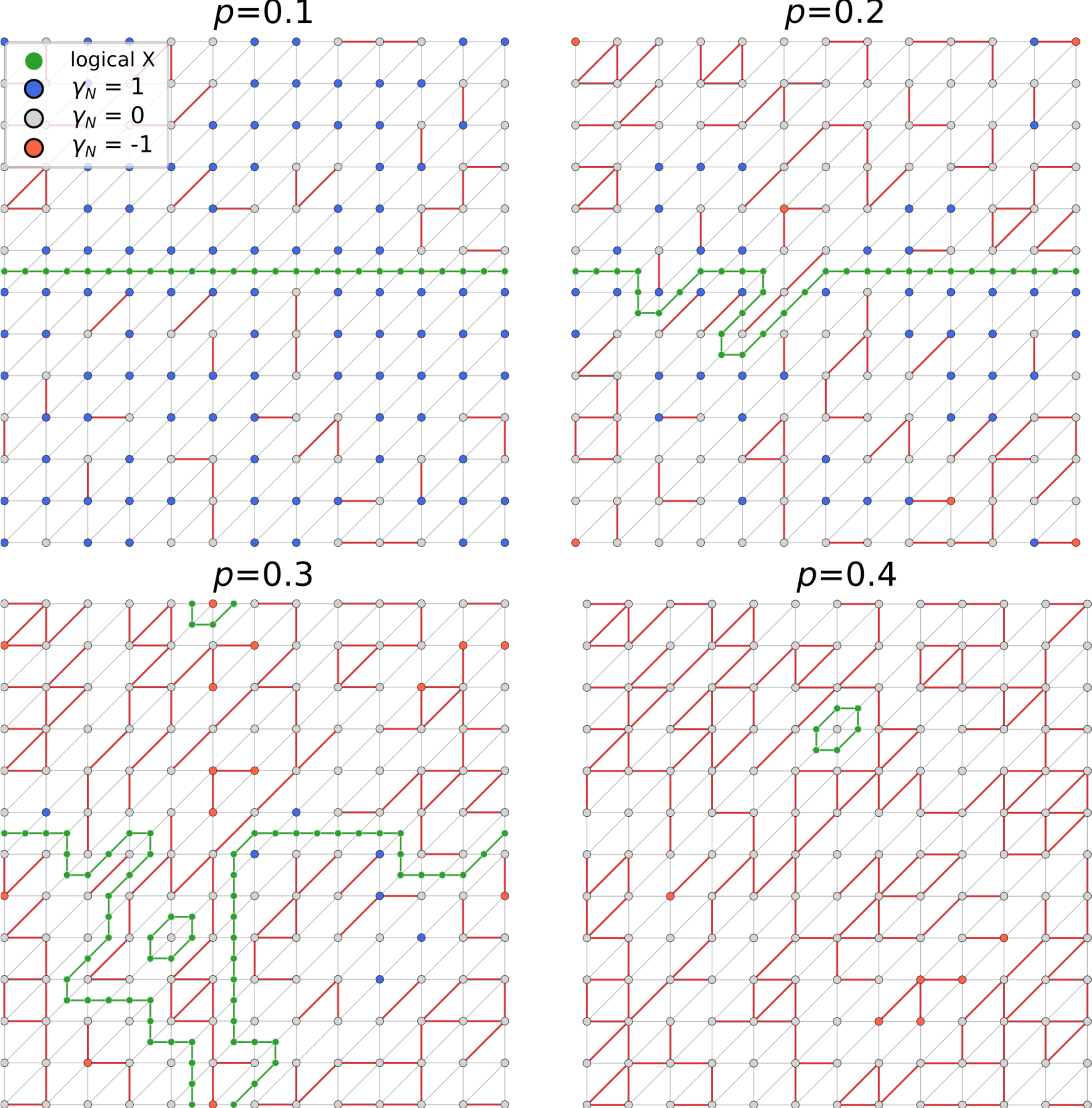}
\end{center}
\caption{Snapshots of EP, QLTEN and surviving logical qubit for various values of $p$ and
$n=1$.
For $p=0.3$, the logical stabilizer deforms and takes a winding path, including a closed loop as a 
part of it.
For $p=0.4$, it disappears.
 }
\label{Fig_TCsnap1}
\end{figure*}
\begin{figure*}[t]
\begin{center}
\includegraphics[width=14cm]{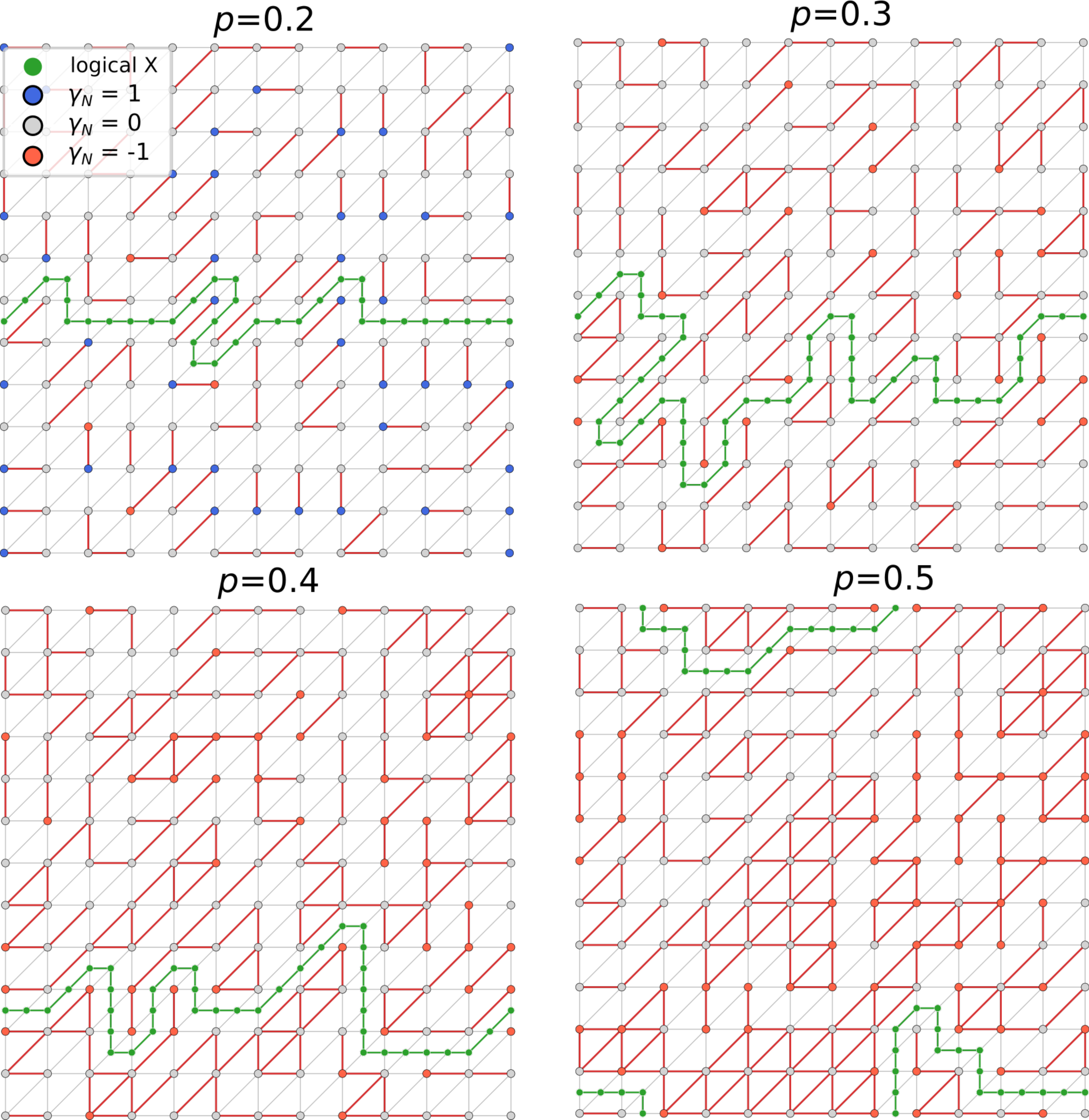}
\end{center}
\caption{Snapshots of EP, QLTEN and surviving logical qubit for various values of $p$ and $n=120$.
Even for $p=0.5$, the logical stabilizer survives.
This behavior correlates with the largest cluster of QLTEN.
}
\label{Fig_TCsnap2}
\end{figure*}

The Hamiltonian of the triangular TC is useful for understanding the properties of the ground state and
the $Z$-decoherence, which is given as follows in terms of Pauli matrices $X_\ell$ and $Z_\ell$ on links $\{\ell\}$,
\be
H_{\rm TC}=-\sum_s A_s -\sum_p B_p, 
\label{HTC1}
\ee
where $A_s= \prod_{\ell\in s} X_\ell$, $\ell\in s$ denotes six links emanating from the site (vertex) $s$, 
and $B_p=\prod_{\ell\in p} Z_\ell$, $\ell\in p$ denotes three links surrounding the triangular plaquette $p$; they are stabilizers.
See Fig.~\ref{Fig_TCH}.
The ground state of $H_{\rm TC}$ satisfies the condition 
$
A_s|\mbox{\rm GS;tc}\rangle=B_p|\mbox{\rm GS;tc}\rangle=|\mbox{\rm GS;tc}\rangle,
$
for all sites and plaquettes.
There are redundancies such as 
$\prod_sA_s=\prod_pB_p=1, $
as we consider the system on a torus.
In order to perfectly specify the ground state, we introduce nonlocal operators defined on 
non-contractible loops wrapping the torus, $\{C_1,C_2\}$ and their counterparts on the dual lattice
$\{\tilde{C}_1,\tilde{C}_2\}$ \cite{PhysRevD.10.2445,thooft1978},
\be
&& W(C_\alpha)=\prod_{\ell \in C_\alpha}Z_\ell,  \nonumber  \\
&& T(\tilde{C}_\alpha)=\prod_{\ell \in \tilde{C}_\alpha}X_\ell,
\ee
where $\alpha=1,2$, and we choose the non-contractible loops such that $W(C_{1(2)})$ and $T(\tilde{C}_{2(1)})$ 
anti-commute each other and otherwise commute.
[In practical calculations, we take $C_{1(2)}=\tilde{C}_{1(2)}$ as in Fig.~\ref{Fig_TCH}.]
In the previous paper \cite{kataoka2026measurement}, we studied the triangular TC under MOC
corresponding to both the measurement of $Z$ and $X$ external fields and clarified the phase diagrams.

In this work, we start with the ground state $|{\rm GS};T(\tilde{C}_{1,2})\rangle$, 
\be
&&T(\tilde{C}_{1})|{\rm GS};T(\tilde{C}_{1,2})\rangle=|{\rm GS};T(\tilde{C}_{1,2})\rangle,\nonumber \\
&&T(\tilde{C}_{2})|{\rm GS};T(\tilde{C}_{1,2})\rangle=|{\rm GS};T(\tilde{C}_{1,2})\rangle,
\label{TCGS}
\ee
and apply $Z$-decoherence on it.
This process is described by the following quantum noisy channel,
\begin{eqnarray}
&&\mathcal{E}^{Z}=\prod_{\ell}\mathcal{E}^{Z}_{\ell},\nonumber\\
&&\mathcal{E}^{Z}_{\ell}[\rho]=(1-p_\ell)\rho+p_\ell Z_\ell\rho Z_\ell,
\label{deco_Z}
\end{eqnarray}
where $p_\ell\in [0,1/2]$ controls the probability of decoherence at the link $\ell$.
As in the previous color code system, we take $p_\ell=0$ or 1/2 by choosing $p_\ell=1/2$ links
with probability $p$ using the $n$-trial EP rule. 
As in the color code case, we shall study the emergent mixed state from $|{\rm GS};T(\tilde{C}_{1,2})\rangle$
mainly using numerical methods in the spirit of network growth of decoherence percolation.
Then, we shall discuss the obtained data focusing on how the logical operators $T(\tilde{C}_\alpha)$
behave as increasing $p$.
As the TC system in an external field in the $Z$-direction being described by
$H_{\rm TC}-h_z\sum_\ell Z_\ell$ exhibits the phase transition to the Higgs phase as the parameter
$h_z$ increases, we call the emergent mixed state Higgs state.

More concretely, the decoherence process in Eq.~(\ref{deco_Z}) is described as follows in the 
stabilizer formalism; By the $Z$-decoherence on the link $\ell$, two stabilizers $A_{s_1}$ and $A_{s_2}$
sharing the link $\ell$ merge into a single stabilizer $(A_{s_1}A_{s_2})$, while $A_{s_1}$, $A_{s_2}$
and $Z_\ell$ are discarded.
Although $A_s$ can be identified with $S^X_{(c,p)}$ of the color code, $A_{s_1}$ and $A_{s_2}$ share
the link $\ell$, and therefore $(A_{s_1}A_{s_2})$ is geometrically different from
$(S^X_{(c,p_1)}S^X_{(c,p_2)})$.
This seemingly subtle difference can cause a substantial distinction in the systems' critical behavior
as we shall see.


\subsection{Numerical study of TEN, large cluster, loop and string operators}

In this subsection, we shall show the results of the numerical calculation of TEN, its variance, etc.
In particular, we are interested in their $n$-dependence, which is to be compared with those of the color code.
In Fig.~\ref{Fig_TCABC} in Appendix A shows the subsystems to calculate TEN and QLTEN.

Figures~\ref{Fig_TC1}(a), (b) show TEN and its variance as a function of $p$ for various types of EP.
The data obviously have clear $n$-dependence, that is, the emergence of the Higgs phase shifts to
larger $p$ as the value of $n$ increases, in particular for the large $n$.
This behavior is related with the appearance of large clusters of the Higgs regime as observed 
in Figs.~\ref{Fig_cluster/TC}, that is, the suppression of large clusters causes a delay in 
the emergence of the Higgs phase.
It is also notable that the QLTEN clusters are strongly correlated with those of EP in Fig.~\ref{Fig_EP}.

Furthermore, for very large $n$, TEN and its variance exhibit a somewhat peculiar behavior, namely,
TEN decreases first below $\gamma_N=0$ and then decreases to $\gamma_N=0$.
As a result, the variance of TEN has a double-peak shape.
This behavior is in sharp contrast to that of the color code and may come from the nature of the evolution process so that, 
as $p$ increases, a large number of moderate size clusters first form, and then these clusters begin to merge into large clusters.
The related behavior of the system is observed by the cluster size distribution in Fig.~\ref{Fig_clustersize}. 
More concrete evidence is obtained by observing the Higgs cluster size distribution of QLTEN, which is 
given in Fig.~\ref{Fig_TCdistribution}.

The above behavior of TEN poses an interesting issue of how the topological order is influenced by 
various types of EP decoherence.
We would like to see how topological entanglement and topological order are inter-related with each other,
and recognize the utility of QLTEN.

To  this end, we study how the non-contractible loop operators, i.e., the logical operators,
evolve under $n$-trial EP decoherence.
As we explained earlier, the initial state is the ground state $|{\rm GS};T(\tilde{C}_{1,2})\rangle$.
By applying $Z$-decoherence, the initial `tHooft loops $T(\tilde{C}_{1(2)})$ deform their shape to 
remain an element of the stabilizer group.
As $p$ increases, a large cluster of $\gamma_N=0$ grows, and the additional application of $Z$-decoherence makes
$T(\tilde{C}_{1(2)})$ unable to stay in the stabilizer group as no other elements of the stabilizers except $T(\tilde{C}_{1(2)})$
anti-commute with the decoherence operation.
At this moment, the target mixed state loses the topological order and ceases to work 
as a logical qubit \cite{botzung2025robustness,kuno2025intrinsic}.
The most interesting issue is how this time step looks from the quantum-entanglement point of view.

In Figs.~\ref{Fig_TC1}(c) and (d), we show the numerical calculations of the expectation values of 
$T(\tilde{C}_{1(2)})$, 
\be
\langle T(\tilde{C}_{1(2)})\rangle \equiv \mbox{Tr}\Big(\rho_{\rm D} T(\tilde{C}_{1(2)})\Big), 
\label{Tloop2}  
\ee
where $\rho_{\rm D}$ is the mixed state that evolves from $|{\rm GS};T(\tilde{C}_{1(2)})\rangle$ under EP decoherence.
In addition to 'tHooft loop (\ref{Tloop2}),
we consider the expectation value of the Wilson string defined as, 
\be 
&& \langle W_{\rm S} \rangle =\frac{1}{L_y}\sum_y W(y),  \nonumber \\
&& W(y)=\frac{1}{L_r}\sum^{L_r}_{r=1} W(y;r), \label{TCWilson} \\
&& W(y;r)=
\mbox{Tr}\Big(\rho_{\rm D} W(\gamma_{y;r}) \rho_{\rm D} W(\gamma_{y;r})\Big)/\mbox{Tr}(\rho_{\rm D}^2), \nonumber
\ee
where $W(\gamma_{y;r})$ is the product of $Z_\ell$ along the string $\gamma_{y;r}$ in Fig.~\ref{Fig_TCH}
of length $r$.
The numerical calculations of the Wilson string $\langle W_{\rm S}\rangle$ are shown 
in Figs.~\ref{Fig_TC1}(e) and (f) for $L_r=11$.

The obtained data obviously show that these observables exhibit a transition behavior to the Higgs phase.
That is, in all cases of $n$, $\langle T(\tilde{C}_{1(2)})\rangle$ maintains unity and suddenly starts to
decrease at some value of $p=p_{\rm T}$, 
while $\langle W_{\rm S} \rangle$ is first vanishingly small and then starts to increase at $p=p_{\rm W}\simeq p_{\rm T}$.
We note that the behavior of the expectation values strongly correlates with that of the largest cluster 
observed by QLTEN as shown in Figs.~\ref{Fig_cluster/TC} for all $n$.
However, we judge that signal of the QLTEN is {\it precocious} to the elimination of the logical qubits, 
namely, the time step at which the logical qubit disappears coincides with the time step
at which the largest cluster of QLTEN expands to the whole system.
This property of QLTEN may propose a new quantum error mitigation by using data of stabilizer syndrome
through QLTEN. 
Simply put, a QLTEN configuration can be constructed through the syndrome by a classical computer,
and it works as an advance warning to destruction of the code space.
We shall report details of this possibility in a future work.

Figures \ref{Fig_TCsnap1} and \ref{Fig_TCsnap2} show some snapshots of the 'tHooft loop 
at time steps in the vicinity of 
its elimination, as well as snapshots of the EP and the QLTEN configurations.
The 'tHooft loop runs avoiding the decohered regions as is qualitatively expected,
although the QLTEN shows that almost of the whole system is already Higgs region $\gamma_N=0$ at the elimination time.
The observations in the above indicate that the QLTEN cluster reacts to decoherence faster than the 1-form symmetry
and the non-contractible logical orders.

The above observation clearly indicates how the quantum entanglement entropy and 1-form symmetry are related to 
each other, as well as how QLTEN is an efficient measure for observing the topological phase transition.
This is one of the main findings of this work.

\section{Conclusion and discussion}

In this work, we studied the topological phase transitions from a microscopic point of view
by using the color code and the TC as case studies.
To this end, we introduced a new measure named QLTEN to describe geometric local configurations of
the intermediate states subject to decoherence.
By observing QLTEN, we elucidated the behavior of the mean TEN to be somehow peculiar from the point of view 
of the continuous phase transition.
Snapshots of QLTEN give us an intuitive picture of the phase transitions and clarify their essential 
relationship to the simplicial homology.
More precisely, our consideration and numerics on 7-plaq QLTEN provide us evidence that TEN is connected to the first homology, 
while the 1-form symmetry is clearly connected to the 0-th homology. 
The close relationship between these two properties is given by the consideration based on QLTEN
and other numerical data, in particular, the largest cluster behavior of the ensemble.

To examine the potential of QLTEN, we have introduced the $n$-trial EP, which plays an important role in the last
two decades.
We verified that QLTEN is faithfully reflected by the properties of EP and the distribution of cluster size
in QLTEN clarifies the geometric property of the topological phases in the TC system, while QLTEN is 
quite insensitive to the nature of EP in the color code.
This discrepancy between the two systems poses a serious problem for universal understanding of the
topological phase transition that remains a subject of future work.

Finally, we studied the relationship between QLTEN and logical qubits in the TC.
By tracing logical qubits on a growing network of decoherence, we identified the time at which
logical qubits disappear, and observed their location just before the elimination.
We found a close relationship between the location of the qubits and the configuration of the EP.
This observation sheds light on the quantum error mitigation of topological quantum computing.
In particular, we emphasize that {\it the microscopic behavior of the topological-changing transition 
in the color code and the TC is rather different from each other}, even though both are promising candidates for
qubits of topological computing and closely related with each other from the anyon point of view.
This issue remains a topic for future work.

We would like to comment on the holomogical property of the transition from the color code to the TC.
As we discussed, QLTEN and the disorder parameter of the 1-form symmetry are closely connected with
the first and the 0-th simplicial homology, respectively although the former was examined
only for the specific type of subsystems used for QLTEN.
There exists an additional aspect of the second homology in that transition.
As explained in Eq.~(\ref{linkqubit}), $XX$-decoherence produces a link qubit on the applied link
and the TC operators, $\check{Z}_\ell$ and $\check{X}_\ell$.
In order to generate a genuine gauge theory on the triangular lattice, the plaquette terms 
$\prod_{\ell \in \triangle} \check{Z}_\ell$ must be generated by decoherence.
The term is nothing but the second homology, and its appearance can be examined by the operator
in Eq.~(\ref{quasiloop}) by setting the loop to $\triangle$, or simply by the decoherence network
configuration generated by Union Find.
In fact from the decoherence data in Fig.~\ref{Fig_TCsnap1} (that is common to the transition from the color code to the TC), 
we can see that even at the time step at which the logical operators
disappear from the stabilizer group ($p=0.4$), the growing network of decoherence is still sparse and,
therefore, the nature of the gauge theory has not been constructed yet.
Anyway, results related to this issue will be reported in future work.

All the findings obtained in this work, in particular those of QLTEN, should be re-examined
by employing different subsystem sizes instead of the 7-plaquette hexagon.
This point is related to the resolution and will be addressed in future work.
In Appendix B, we display some of the numerical results corresponding to this issue,
i.e., behavior of clusters observed through the QLTEN using 12-plaquette subsystems.
Data show that for both the color code and toric code, the cluster systems exhibit similar behavior to those
probed by the 7-plaquette hexagons, indicating the stability of the QLTEN at least for relatively small
subsystem probes.

Finally, we would like to give a brief comment on the recently proposed idea, the entanglement asymmetry~\cite{ares2023entanglement}.
The entanglement asymmetry is a subsystem measure of symmetry breaking, focusing on the reduced density matrix of 
a subsystem $\rho_A$ and its symmetrized version with respect to the target symmetry $\rho_{AS}$.
By measuring the distance between two density matrices by relative entropy, information is obtained about how much 
a symmetry is broken.
This idea was applied to 1-form symmetry~\cite{benini2025entanglement,lamas2025higher,benini2026higher}, 
and it was discussed how the 1-form symmetries are realized in sub-regions of the system.
As we consider the subsystem entanglement entropy in this work, it is an interesting future work to 
consider the hybrid of these ideas.

Another interesting direction of search is the application of the present percolation perspective to
non-Hermitian physics.
In this vein, in Ref.~\cite{yang2023percolation}, the topological properties of non-Hermitian systems are
investigated.




\section*{Data availability}
Data supporting the findings of this study are available from the authors on reasonable request.\\

\bigskip
{\it Acknowledgments} \\
The authors thank Takahiro Orito at Nihon University for his useful advice on the numerical study of the loop operators.
The computation in this work has been done using the facilities of the Supercomputer Center, the Institute for Solid State Physics, the University of Tokyo (ISSPkyodo-SC-2026-Ba-0068).



\onecolumngrid

\renewcommand{\thesection}{A\arabic{section}} 
\renewcommand{\theequation}{A\arabic{equation}}
\renewcommand{\thefigure}{A\arabic{figure}}
\setcounter{equation}{0}
\setcounter{figure}{0}
\appendix

\setcounter{section}{0}
\renewcommand{\thesection}{\Alph{section}}

\makeatletter
\@addtoreset{figure}{section}
\makeatother
\renewcommand{\thefigure}{\Alph{section}\arabic{figure}}

\makeatletter
\renewcommand{\theHsection}{\Alph{section}}
\renewcommand{\theHfigure}{\thesection.\arabic{figure}}
\renewcommand{\theHtable}{\thesection.\arabic{table}}
\renewcommand{\theHequation}{\thesection.\arabic{equation}}
\makeatother


\section{Additional schema of the models and the finite-size scaling analysis}

In this appendix, we give some details of the methods to calculate TEN and QLTEN,
the finite-size scaling analysis, and the results obtained.

\subsection{Prescription of TEN and QLTEN}

We show the protocol for calculating TEN and QLTEN.
As explained in the main text, we have to specify the shapes of the subsystems $A,B.C$.
Essential points are the same for the color code and the TC, whereas the color code (TC)
is defined on the honeycomb (triangular) lattice.

\begin{figure*}[t]
\begin{center}
\centering
\includegraphics[width=12cm]{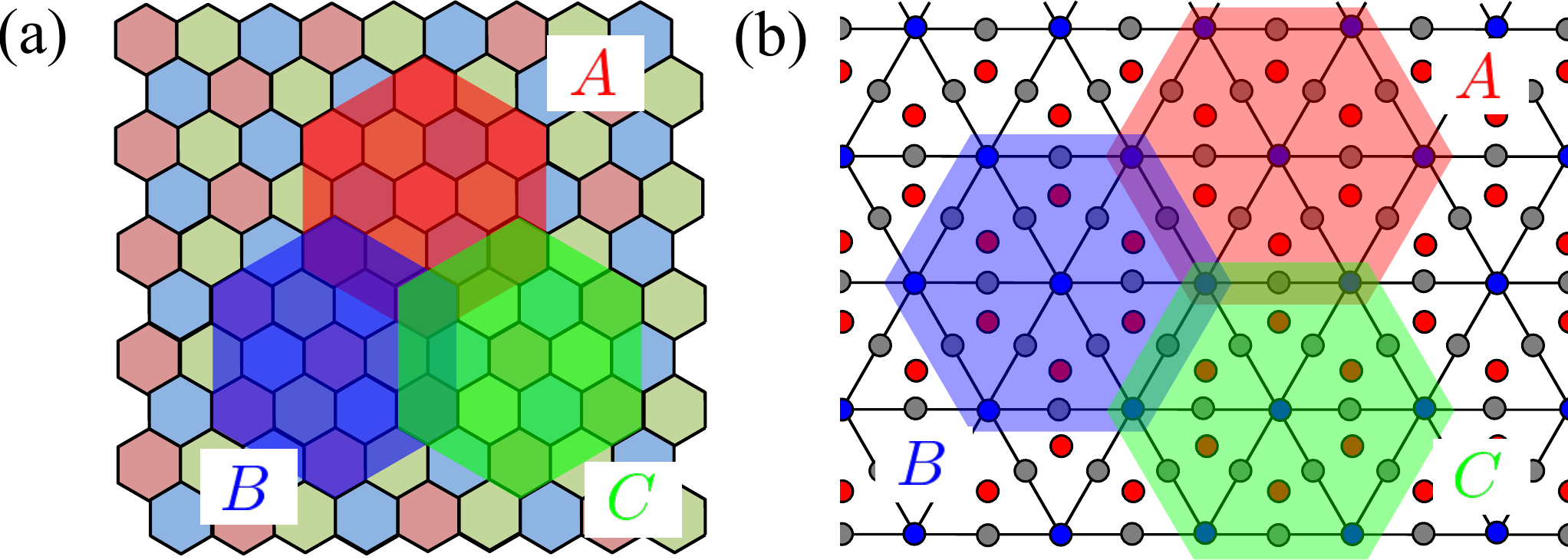}
\end{center}
\caption{Samples of subsystems to calculate TEN and QLTEN in the color code (a) and toric code (b). 
}
\label{Fig_TCABC}
\end{figure*}

The first step is to construct a hexagonal complex by composing the plaquette stabilizers,
which is an essential ingredient for the calculation of TEN and QLTEN in both the color code and the TC.
The hexagonal complexes and their relative locations are shown in Fig.~\ref{Fig_TCABC}, in which
there is a slight difference between the color code and the TC, whereas from the point of view 
of the red triangular lattice of the color code, there exists a clear correspondence
between the color code and the TC.

In the color code, this type of hexagon is called 7-plaquette hexagon.
The larger subsystems used for the calculation are obtained by expanding and merging the hexagons
in both the color code and the TC.
In the previous paper~\cite{kataoka2026measurement}, we showed that the way using hexagons provides us
with stable numerical results without ambiguity.

\subsection{the finite-size scaling analysis}
In this subsection, we explain the finite-size scaling (FSS) analysis
for the string data in Figs.~\ref{Fig_TEN1} and \ref{Fig_largen}~\cite{kuno2025intrinsic,kataoka2026measurement}.
We employ the following Ansatz,
\be
D^X(\Gamma)=\ell ^ {-\zeta} F((p-p_c) \ell ^{1 / \nu}),
\label{FSS1}
\ee
where $\ell=|\Gamma|=\mbox{length of} \;\Gamma$, 
$F$ is a scaling function and $\nu$ is the critical exponent.
We tried to apply FSS to the variance of $D^X(\Gamma)$, but could not obtain 
satisfactory results.
Then, we show the FSS for $D^X(\Gamma)$ that produces reliable results.

\begin{figure*}[htbp]
\begin{center}
\centering
\includegraphics[width=13cm]{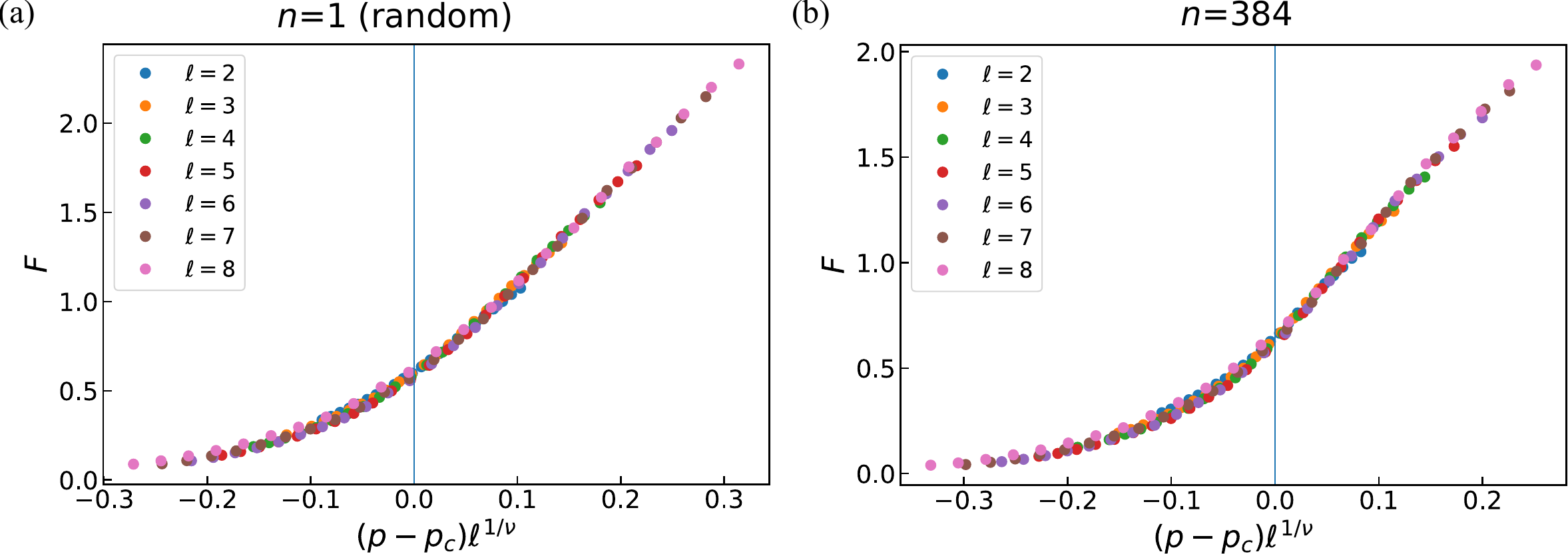}
\end{center}
\caption{Finite-size scaling of $D^X(\Gamma)$ for $n=1$ and $n=384$.
All data collapse into a single curve. 
}
\label{Fig_FSS}
\end{figure*}

In Fig.~\ref{Fig_FSS}, we show the scaling functions of $D^X(\Gamma)$ obtained by the FSS.
All data collapse into a single curve and we estimate $p_s$ and $\nu$ using Eq.~(\ref{FSS1}).
The values obtained are shown in table \ref{Table1}.


\begin{table*}[thbp]
\centering
\caption{Table IA:
Values of the critical point $p_s$ and critical exponent $\nu$ for $n=1$, obtained from finite-size scaling analyses with different minimum string lengths $\ell_{\min}$, while fixing the maximum string length at $\ell_{\max}=8$.
Table IB: Values of $p_s$ and $\nu$ for $n=384$, obtained using the same procedure.
}
\label{Tab1}
\begin{tabular}{ccc@{\hspace{7.5mm}}c@{\hspace{7.5mm}}ccc}
    \multicolumn{3}{c}{Table IA}
    & 
    & \multicolumn{3}{c}{Table IB} \\
    \multicolumn{3}{c}{$n=1$}
    &
    & \multicolumn{3}{c}{$n=384$} \\
    \cline{1-3}\cline{5-7}
    $\ell_{\min}$ & $p_s$ & $\nu$
    &
    & $\ell_{\min}$ & $p_s$ & $\nu$ \\
    \cline{1-3}\cline{5-7}
    1 & $0.314 \pm 0.003$ & $1.24 \pm 0.02$
    &
    & 1 & $0.320 \pm 0.002$ & $1.19 \pm 0.02$ \\
    2 & $0.301 \pm 0.003$ & $1.24 \pm 0.03$
    &
    & 2 & $0.312 \pm 0.002$ & $1.25 \pm 0.03$ \\
    3 & $0.298 \pm 0.005$ & $1.30 \pm 0.06$
    & 
    & 3 & $0.314 \pm 0.003$ & $1.36 \pm 0.04$ \\
    \cline{1-3}\cline{5-7}
\end{tabular}
\label{Table1}
\end{table*}

\section{Quasi-local entanglement negativity observed by 12-plaq subsystems}

\begin{figure*}[htbp]
\begin{center}
\centering
\includegraphics[width=7.7cm]{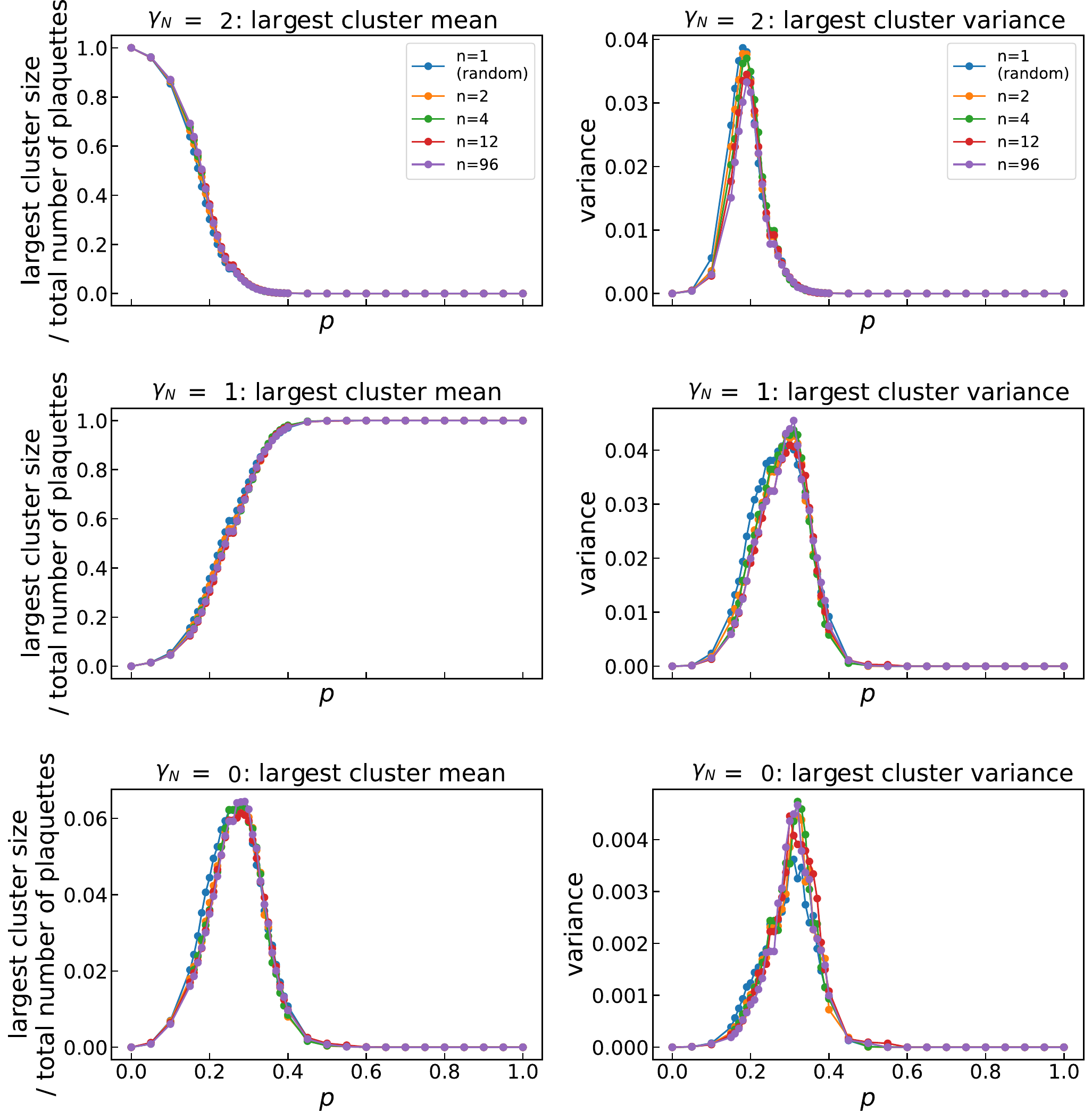} \hspace{0.5cm}
\includegraphics[width=8.3cm]{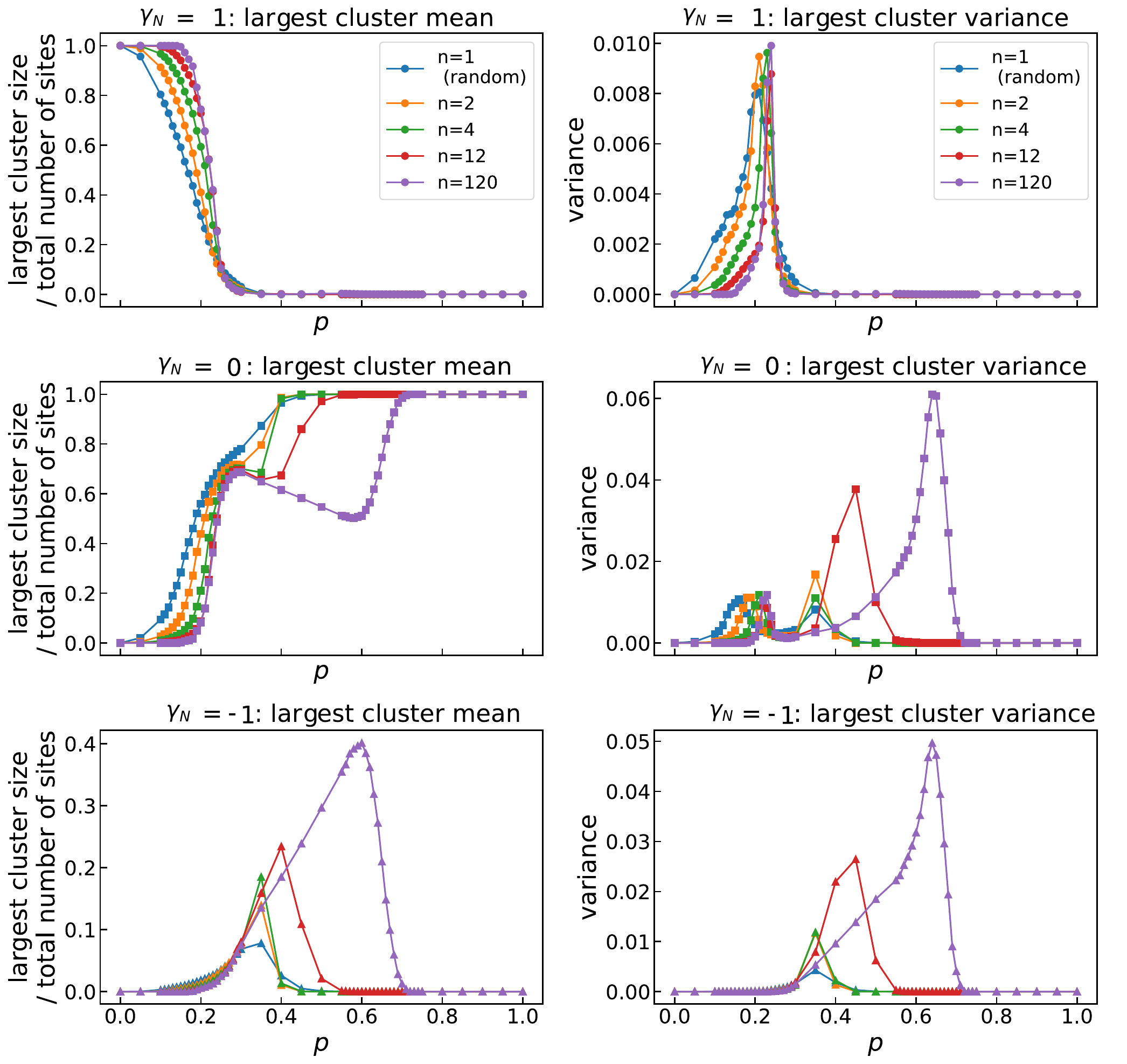}
\end{center}
\caption{Left: Behavior of clusters in the color code calculated through 12-plaquette subsystems.
Right: Behavior of clusters in the toric code calculated through 12-plaquette subsystems.
These data should be compared with those in Figs. \ref{Fig_CCcluster} and \ref{Fig_cluster/TC}, respectively.
It is observed that the clusters observed through two kinds of subsystems exhibit qualitatively 
similar behaviors, although the variances get larger for larger subsystem.
}
\label{Fig_cluster2}
\end{figure*}

In this appendix, we perform the numerical calculation of the QLTEN observed by 12-plaq subsystem probe.
Obtained data are shown in Fig. \ref{Fig_cluster2}, which are to be compared with those in Figs. \ref{Fig_CCcluster} and \ref{Fig_cluster/TC}.
It is clearly observed that for both the color code and toric code, cluster systems behave quite similarly
to those observed by 7-plaq hexagon subsystems.
Therefore, the stability of QLTEN is verified for at least relatively small subsystem probes.
Study on the behavior of QLTEN observed by using larger subsystems is a future issue.




\newpage
\twocolumngrid
\FloatBarrier
{
\bibliography{ref2}

\end{document}